\documentclass[twoside,twocolumn,9pt]{article}
\usepackage{extsizes}
\usepackage[super,sort&compress,comma]{natbib} 
\usepackage[version=3]{mhchem}
\usepackage[left=1.5cm, right=1.5cm, top=1.785cm, bottom=2.0cm]{geometry}
\usepackage{balance}
\usepackage{mathptmx}
\usepackage{sectsty}
\usepackage{graphicx} 
\usepackage{lastpage}
\usepackage[format=plain,justification=justified,singlelinecheck=false,font={stretch=1.125,small,sf},labelfont=bf,labelsep=space]{caption}
\usepackage{float}
\usepackage{fancyhdr}
\usepackage{fnpos}
\usepackage[english]{babel}
\addto{\captionsenglish}{%
  
}
\usepackage{array}
\usepackage{droidsans}
\usepackage{charter}
\usepackage[T1]{fontenc}
\usepackage[usenames,dvipsnames]{xcolor}
\usepackage{setspace}
\usepackage[compact]{titlesec}
\usepackage{hyperref}

\usepackage[normalem]{ulem}

\usepackage{epstopdf}

\definecolor{cream}{RGB}{222,217,201}

\begin{document}

\pagestyle{fancy}
\thispagestyle{plain}
\fancypagestyle{plain}{
\renewcommand{\headrulewidth}{0pt}
}

\makeFNbottom
\makeatletter
\renewcommand\LARGE{\@setfontsize\LARGE{15pt}{17}}
\renewcommand\Large{\@setfontsize\Large{12pt}{14}}
\renewcommand\large{\@setfontsize\large{10pt}{12}}
\renewcommand\footnotesize{\@setfontsize\footnotesize{7pt}{10}}
\makeatother

\renewcommand{\thefootnote}{\fnsymbol{footnote}}
\renewcommand\footnoterule{\vspace*{1pt}%
\color{cream}\hrule width 3.5in height 0.4pt \color{black}\vspace*{5pt}} 
\setcounter{secnumdepth}{5}

\makeatletter 
\renewcommand\@biblabel[1]{#1}            
\renewcommand\@makefntext[1]%
{\noindent\makebox[0pt][r]{\@thefnmark\,}#1}
\makeatother 
\renewcommand{\figurename}{\small{Fig.}~}
\sectionfont{\sffamily\Large}
\subsectionfont{\normalsize}
\subsubsectionfont{\bf}
\setstretch{1.125} 
\setlength{\skip\footins}{0.8cm}
\setlength{\footnotesep}{0.25cm}
\setlength{\jot}{10pt}
\titlespacing*{\section}{0pt}{4pt}{4pt}
\titlespacing*{\subsection}{0pt}{15pt}{1pt}

\fancyfoot{}
\fancyfoot[LO,RE]{\vspace{-7.1pt}\includegraphics[height=9pt]{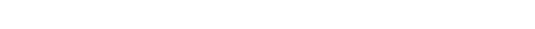}}
\fancyfoot[CO]{\vspace{-7.1pt}\hspace{11.9cm}\includegraphics{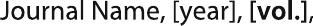}}
\fancyfoot[CE]{\vspace{-7.2pt}\hspace{-13.2cm}\includegraphics{head_foot/RF}}
\fancyfoot[RO]{\footnotesize{\sffamily{1--\pageref{LastPage} ~\textbar  \hspace{2pt}\thepage}}}
\fancyfoot[LE]{\footnotesize{\sffamily{\thepage~\textbar\hspace{4.65cm} 1--\pageref{LastPage}}}}
\fancyhead{}
\renewcommand{\headrulewidth}{0pt} 
\renewcommand{\footrulewidth}{0pt}
\setlength{\arrayrulewidth}{1pt}
\setlength{\columnsep}{6.5mm}
\setlength\bibsep{1pt}

\makeatletter 
\newlength{\figrulesep} 
\setlength{\figrulesep}{0.5\textfloatsep} 

\newcommand{\topfigrule}{\vspace*{-1pt}%
\noindent{\color{cream}\rule[-\figrulesep]{\columnwidth}{1.5pt}} }

\newcommand{\botfigrule}{\vspace*{-2pt}%
\noindent{\color{cream}\rule[\figrulesep]{\columnwidth}{1.5pt}} }

\newcommand{\dblfigrule}{\vspace*{-1pt}%
\noindent{\color{cream}\rule[-\figrulesep]{\textwidth}{1.5pt}} }

\makeatother

\twocolumn[
  \begin{@twocolumnfalse}
{\includegraphics[height=30pt]{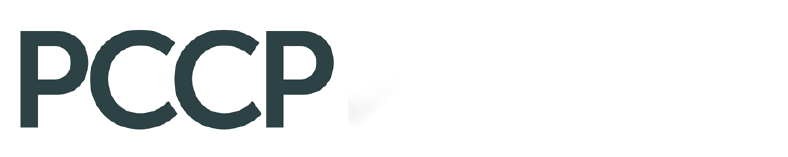}\hfill\raisebox{0pt}[0pt][0pt]{\includegraphics[height=55pt]{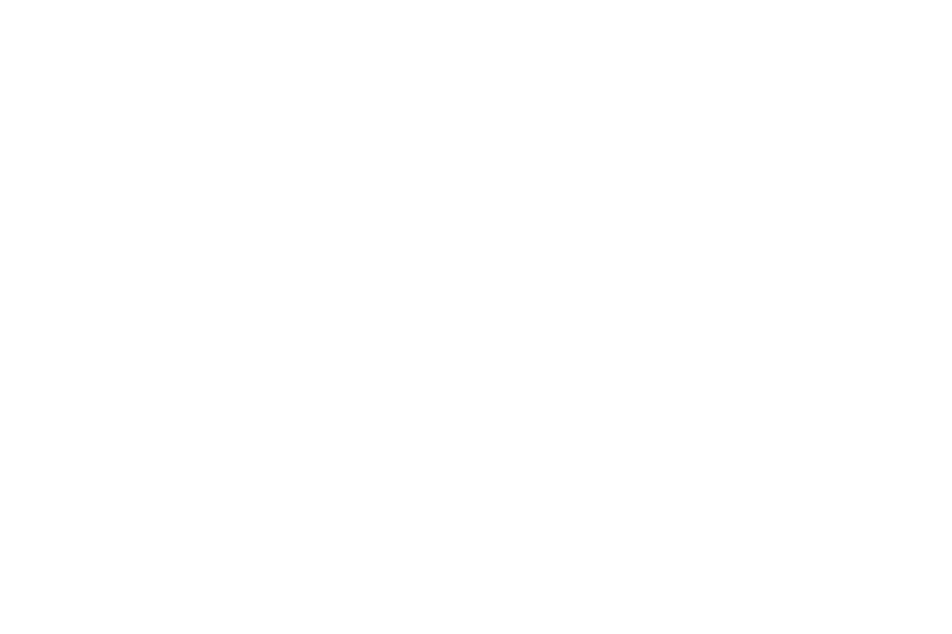}}\\[1ex]
\includegraphics[width=18.5cm]{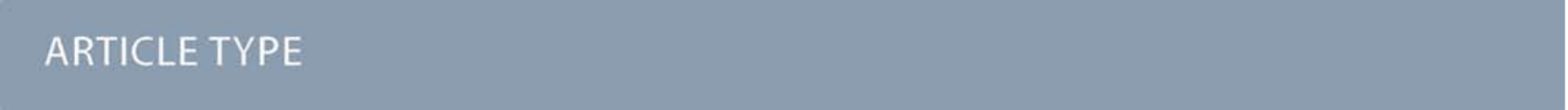}}\par
\vspace{1em}
\sffamily
\begin{tabular}{m{4.5cm} p{13.5cm} }

\includegraphics{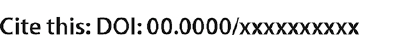} & \noindent\LARGE{\textbf{A systematic study of the valence electronic structure of cyclo(Gly-Phe), cyclo(Trp-Tyr) and cyclo(Trp-Trp) dipeptides in gas phase$^\dag$  } } \\
\vspace{0.3cm} & \vspace{0.3cm} \\

 & \noindent\large{ Elena Molteni,$^{\ast}$\textit{$^{a,b}$} Giuseppe Mattioli,\textit{$^{a}$} Paola Alippi,\textit{$^{a}$} Lorenzo Avaldi,\textit{$^{a}$} Paola Bolognesi,\textit{$^{a}$} Laura Carlini,\textit{$^{a}$} Federico Vismarra,\textit{$^{c1,c2}$} Yingxuan Wu,\textit{$^{c1,c2}$} Rocio Borrego Varillas,\textit{$^{c2}$} Mauro Nisoli,\textit{$^{c1,c2}$} Manjot Singh,\textit{$^{d}$} Mohammadhassan Valadan,\textit{$^{d,e}$}  Carlo Altucci,\textit{$^{d,e}$}  Robert Richter\textit{$^{f}$} and Davide Sangalli\textit{$^{a,b}$} } \\

\includegraphics{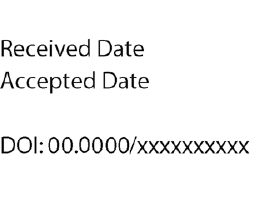} & \noindent\normalsize{The electronic energy levels of cyclo(Glycine-Phenylalanine), cyclo(Tryptophan-Tyrosine) and cyclo(Tryptophan-Tryptophan)  dipeptides are investigated with a joint experimental and theoretical approach.
Experimentally, valence photoelectron spectra in the gas phase are measured using VUV radiation.
Theoretically, we first obtain low-energy conformers through an automated conformer-rotamer ensemble sampling scheme based on tight-binding simulations. Then, different first principles computational schemes are considered to simulate the spectra: Hartree-Fock (HF), density functional theory (DFT) within the B3LYP approximation, the quasi--particle GW correction, and the quantum-chemistry CCSD method. Theory allows to assign the main features of the spectra. A discussion on the role of electronic correlation is provided, by comparing computationally cheaper DFT scheme (and GW) results with the accurate CCSD method.} \\

\end{tabular}

 \end{@twocolumnfalse} \vspace{0.6cm}

  ]

\renewcommand*\rmdefault{bch}\normalfont\upshape
\rmfamily
\section*{}
\vspace{-1cm}


\footnotetext{\textit{$^{a}${Istituto di Struttura della Materia-CNR (ISM-CNR), Area della Ricerca di Roma 1, Via Salaria km 29,300, CP 10, Monterotondo Scalo, Roma, Italy; E-mail (EM): elena.molteni@mlib.ism.cnr.it}}}
\footnotetext{\textit{$^{b}${Dipartimento di Fisica, Università degli Studi di Milano, via Celoria 16, I-20133 Milano, Italy}}}
\footnotetext{\textit{$^{c1}${Dipartimento di Fisica, Politecnico di Milano, Piazza Leonardo da Vinci, 32, Milano, Italy}}}
\footnotetext{\textit{$^{c2}${CNR-Istituto di Fotonica e Nanotecnologie, Piazza Leonardo da Vinci, 32, Milano, Italy}}}
\footnotetext{\textit{$^{d}${Dipartimento di Scienze Biomediche Avanzate, Università degli Studi di Napoli Federico II, via Pansini 5, I-80131 Napoli, Italy}}}
\footnotetext{ \textit{$^{e}${Istituto Nazionale Fisica Nucleare (INFN), Sezione di Napoli, Napoli, Italy} }} 
\footnotetext{\textit{  $^{f}$ {Sincrotrone Trieste, Area Science Park,  Basovizza, Trieste, Italy}}}

\footnotetext{\dag~Electronic Supplementary Information (ESI) available: Text and figures about unbound electronic states in the continuum: Figure S1: calculated DOS for both occupied and empty electronic states of c-GlyPhe, obtained with different computational methods;
Figure S2: spatial localization of "resonance" states and of completely delocalized states. 
Text and tables (Tables S1, S2, S3, S4) on the energy ordering of c-TrpTrp conformers and its dependence on several computational details. 
Figures S3 ans S4: Spatial localization of EOM-CCSD and of B3LYP orbitals of c-GlyPhe.
Tables S5 and S6: calculated vertical and adiabatic ionization energies and electron affinities for the cyclo-dipeptides under study.
See DOI: 10.1039/cXCP00000x/}


\section{Introduction}
Peptides are short chains of amino acids covalently linked by peptide bonds between the carboxylic terminal group of each amino acid and the amino terminal group of the following one. They are said to be cyclic if they contain a circular sequence of bonds.   
Cyclo-dipeptides (CDPs) or 2,5-diketopiperazines (DKP), derived from the cyclization of a dipeptide, {\it i.e.}, of a peptide made by two amino acids, are the smallest cyclic peptides: they are widely present in nature and they display a variety of biological and pharmacological activities (such as antibacterial, antiviral, antitumoral, antioxidant).\cite{Mishra_Molecules_2017} They have higher structural rigidity and enzymatic stability with respect to linear peptides, and they feature multiple hydrogen bonding sites, which can potentially have a role both in self-assembly\cite{Mattioli_SciRep_2020} and in forming complexes with different molecules, {\it e.g.} proteins.\cite{Zhao_PepSci_2021}
CDPs and their derivatives have therefore received attention both in drug discovery and as possible building blocks for nanodevices. \cite{Jeziorna_CrystGrowthDes_2015,Zhao_PepSci_2021} 

DKPs can be considered as precursors of longer oligopeptides, they have been detected, {\it e.g.}, in meteorites,~\cite{Danger_ChemSocRev_2012} and, as chiral molecules, they can catalyze enantioselective reactions.~\cite{Ying_SciRep_2018}  
They may therefore have had a role in the synthesis of the first biomolecules on Earth, and in the emerging of the so-called ``homochirality of life'', {\it i.e.} the fact that naturally occurring proteins are composed of L-amino acids.~\cite{Danger_ChemSocRev_2012}
The formation of cyclo-dipeptides in simulated prebiotic conditions has been studied in Ref.~\cite{Ying_SciRep_2018}.
The correlation between chirality and self-organization of cyclo-dipeptides has also been investigated in the literature.~\cite{Jeziorna_CrystGrowthDes_2015} 

Biomolecules are usually studied in solution, often in water, in order to mimic physiological conditions. The presence of the solvent can significantly affect their chemical and physical properties and indeed the analysis of the effects of type of solvent, pH and temperature on {\it e.g.} protein stability and conformation is an interesting field of research.
On the other hand, in spite of the challenge to evaporate these molecules intact, gas phase characterization opens the way to determining the intrinsic properties of the molecule without environmental interferences. It also allows to investigate fundamental processes like ultra-fast charge migration via pump and probe experiments.\cite{Calegari_Science2014} Photoinduced charge transfer through molecules is involved in a variety of mechanisms and phenomena, including charge separation in photosynthetic centers, charge generation in organic solar cells, photoprotection and photodamage.\cite{Nisoli_ChemRev_2017,Krausz_RevmodPhys_2009,Gao_PCCP_2014,Weinkauf_JPC_1996,Rozzi_JPhysCondMat_2018}

Experimental works on cyclo-dipeptides or linear dipeptides have already been reported in the literature, {\it e.g.}, the study of the correlation between photoelectron spectra and structures in c-PhePhe, c-TyrPro and c-HisGly \cite{Arachchilage_JCP_2012} or in c-GlyGly, c-PhePro and c-LeuPro,\cite{Arachchilage_JCP_2010} the laser UV photoionization of the linear TrpTyr and TrpTrp dipeptides\cite{Pileni_ChemPhysLett_1977} and the investigation on aromatic cyclo-dipeptides, such as c-TrpTrp, in methanol solution as building blocks for self-assembled biocompatible supramolecular structures with photoluminescence properties.\cite{Tao_NatComm_2018} The electron mobility and dissociation of several different short linear peptides in the gas phase via UV and IR multiphoton absorption,\cite{Weinkauf_JPC_1995} as well as the fragmentation of c-AlaAla, a process relevant to the understanding of the early stages of evolution of life,\cite{Barreiro-Lage_JPCL2021} have attracted some interest.

Here a joint experimental and theoretical study of the electronic properties of c-GlyPhe, c-TrpTyr, and c-TrpTrp has been performed. The valence photoemission spectra (PES) in gas phase have been measured using synchrotron radiation and interpreted with the help of {\it ab-initio} simulations.
In our computational protocol, stable geometries of DKP in gas phase are extracted through a large-scale conformational search carried out using tight-binding (TB) simulations.
Density functional theory (DFT) calculations are then used to compute the gas phase electronic properties of the most stable conformers. The electronic energy levels and the resulting densities of states (DOS) are compared with the measured PE spectra. 

\section{Methods}

\subsection{Theory}

\subsubsection{Energy minimization and structural relaxation.}
The GFN2-xTB tight-binding Hamiltonian\cite{Grimme_JCTC_2019} has been used as “engine” for the search of minimum energy configurations through an automated conformer–rotamer ensemble sampling tool (CREST) code.\cite{Grimme_JCTC_2019,Grimme_JCTC_b_2019,Mattioli_SciRep_2020,Mattioli_PCCP_2021} The code provides the coordinates of a series of conformers of the molecule, ordered according to their increasing (/decreasing) energy (/Boltzmann population). \\
For each of the three dipeptides, a few lowest energy conformers found with this method were then subjected to a geometry optimization run, and, subsequently, to self-consistent electronic structure calculations: both calculations were performed within DFT~\cite{DFT_HK_PR1964,DFT_KS_PR1965} using either the Quantum ESPRESSO (QE) ~\cite{QE_2017,QE_JPhysCondMat2009}  or the ORCA \cite{Orca_general} suites of programs. There is a generally close agreement on the energy ordering of molecular configurations between tight-binding and {\it ab initio} calculations. A very detailed investigation on this point has been performed in the case of c-TrpTrp, and is discussed in the Supplementary Information.\dag

In the case of QE, simulations of the same molecules have been performed in the framework of a plane-waves/pseudopotential method. Initial structures selected by the CREST algorithm have been accommodated in large cubic supercells (40 a.u., {\it i.e.} $\approx$ 21.167 \AA, side length) and fully optimized. Then, in view of the more computationally demanding GW calculations, also a face-centered cubic (FCC) cell has been used, yielding the same distance between molecules in replicated cells with a smaller volume and, therefore, a lower number of G-vectors.
Total energies have been calculated using norm-conserving Troullier-Martins atomic pseudopotentials,\cite{TM_PsP} a plane-wave basis set, and the B3LYP hybrid exchange-correlation functional,\cite{B3LYP1,B3LYP2} (and also the M062X functional\cite{ref_M062X} in some cases) including the D3 pairwise dispersion correction for van der Waals (vdW) interactions.\cite{Grimme_D3}
Satisfactorily converged results have been achieved by using cutoffs of 90 Ry on the plane waves and of 360 Ry on the electronic density, respectively. 
H 1s$^2$ (2 electrons), C 2s$^2$-2p$^2$ (4 electrons), N 2s$^2$-2p$^3$ (5 electrons), and O 2s$^2$-2p$^4$ (6 electrons) electrons have been treated as valence electrons. All of the inner shell electrons are embedded in the pseudopotentials. The Makov-Payne correction to the
total energy was also computed and an estimate of the vacuum
level was calculated, which allowed us to properly align electronic energy levels.~\cite{MakovPayne} For geometry relaxation runs the Broyden-Fletcher-Goldfarb-Shanno (BFGS) scheme was used.~\cite{BFGS}

In the case of ORCA, DFT simulations have been performed in an all-electron localized-basis-set framework.\cite{Orca_general,Orca_WIREs_2018} Kohn-Sham orbitals have been expanded on large, Gaussian-type def2-QZVPP basis sets.\cite{Ahlrichs_JCP_1992,Ahlrichs_PCCP_2005} The corresponding def2/J basis has been also used as an auxiliary basis set for Coulomb fitting in a resolution-of-identity/chain-of-spheres (RIJCOSX) framework. Molecular geometries have been fully re-optimized and their properties investigated by using the same B3LYP functional used in the case of QE, including the D3 pairwise dispersion correction.

\subsubsection{Energy levels.}
We then analyzed the electronic levels obtained from the B3LYP simulations, with the aim of describing the photoemission measurements. Even if DFT is in principle suitable for reproducing charged excitations using orbital-dependent Koopman's compliant functionals,\cite{Marzari_PRB_2014_functionals} the use of the B3LYP eigenvalues and wave-functions to model photoemission is formally not justified. 
In particular for empty orbitals, which are not even affected by the DFT constraint of reproducing the exact density.
On the other hand, such an approach can be generally justified {\it a posteriori} by the close similarity often obtained between B3LYP energies and photoemission measurements, as well as by comparison with accurate yet much more expensive calculations. We proceed therefore to a detailed comparison with other approaches, using GlyPhe in particular as a test case. For GlyPhe only, we compare the energy levels calculated at the B3LYP level against Hartree-Fock (HF) ones, also obtained via a self consistent field (SCF) calculation done with QE.
Moreover, for the first ionization energy (IE) and the electron affinity (EA) we compute total energy differences using the B3LYP functional and also the M062X functional.
Validation of a DFT-based approach is a prerequisite to use such results as starting points for the subsequent application of many body perturbation theory (MBPT) methods; B3LYP eigenvalues are then used as a starting point for the quasiparticle (QP) corrections calculated within the GW approximation with the Yambo code.~\cite{Sangalli2019} The HF solution is instead the starting point for an additional calculation with the Equation of Motion (EOM) Coupled Cluster Single and Double excitations method (CCSD) with the ORCA code. Both EOM-CCSD and GW give a formal description of photoemission. CCSD directly gives an approximation to the photoemission [and inverse photoemission] energies $(E^N_0-E^{N-1}_i)$ [and $(E^{N+1}_i-E^N_0)$] and wave--functions including dynamical correlations on top of HF.
Here $E^N_0$ is the total energy of the ground state in presence of $N$ electrons, while $E^{N+1}_i$ and $E^{N-1}_i$ are the total energies of excited states with $N\pm1$ electrons.
GW describes the same energies within the quasi--particle approximation  $\epsilon_i\approx (E^N_0-E^{N-1}_i)$ for $\epsilon_i<\mu$ [and $\epsilon_i\approx (E^{N+1}_i-E^N_0)$ for $\epsilon_i>\mu$], with $\mu$ the chemical potential, and assuming that the Kohn--Sham (KS) wave--functions are a good approximation to QP wave--functions.~\cite{Onida2002,Golze2019} Both EOM-CCSD and GW calculations give a renormalization of the peak intensity due to coupling with either independent--particles (IP) transitions, in the CCSD case, or correlated Random Phase Approximation (RPA) transitions or plasmons, in the GW case. EOM-CCSD simulations are performed for the lowest energy conformer of each dipeptide, while GW calculations only for the lowest energy conformer of GlyPhe, for which we also directly compute $(E^N_0-E^{N-1}_0)$ and $(E^{N+1}_0-E^N_0)$, both using the B3LYP and the M062X functional, performing in each of the two cases three different total energy simulations changing the number of electrons in the system.
For the GW scheme we compute the RPA screening within the Plasmon Pole Approximation (PPA), with 400 bands and an energy cutoff of 2 Ha. The GW self--energy is then constructed using 60 bands in the GW summation. 

Reference values for molecular ionization energies have been obtained using ORCA, in the same gaussian-type orbitals (GTO) framework discussed above but through a different wavefunction approach. As introduced above, the simulations are based on the equation of motion - coupled cluster theory (EOM-CC), which extends a primarily ground-state method like the single-reference coupled-cluster one to the calculation of excited/ionized states as well as to the corresponding excitation/ionization energies.\cite{Bartlett_WIREs_2012} Within the ORCA implementation, core and valence ionization energies, as well as electron attachment energies for the mapping of unoccupied electronic states, are calculated by grouping electron pairs in domain-local pairs of natural orbitals (DLPNO), and the chain-of-sphere approximation is used to speed up the calculation of exchange-like integrals with four virtual labels (COSX). \cite{Dutta_JCP_2016,Dutta_JCP_2016_b} The present implementation makes the treatment of large systems (2196 basis functions in the case of the largest c-TrpTrp) possible, with a loss of accuracy lower than 0.01 eV with respect to canonical algorithms.

\subsection{Experimental}
The measurements of the photoelectron spectra have been performed at the CiPo beamline ~\cite{CIPO_1995} of the Elettra synchrotron facility in Trieste, Italy. The radiation from the electromagnetic elliptical wiggler insertion device ~\cite{Walker_1992} in section 4.2 of the storage ring is monochromatized by a variable angle spherical grating monochromator ~\cite{Padmore_1989} and focused by a toroidal mirror to the interaction region of the end station equipped by a VG-220i hemispherical electron energy analyser.~\cite{Plekan_2020}
The hemispherical electron energy analyser is mounted in the plane perpendicular to the direction of propagation of the photon beam, at the magic angle with respect to the electric vector, to avoid that the intensities of the different features in the photoelectron spectrum are affected by the asymmetry parameters of the ionized states.
The hemispherical analyser is equipped with a 2D position sensitive detector\cite{Cautero_2008} which spans a kinetic energy range of about 10\% of the analyser pass energy and is characterized by an energy resolution of about 2\% of the pass energy. In the present experiment a photon energy of 60 eV and an analyser pass energy of 30 eV have been used, resulting in an overall energy resolution of about 600 meV. The electrons kinetic energy scale and energy resolution have been calibrated against the He 1s peak and confirmed by the photoemission peak of water molecules present as background gas in the vacuum chamber or desorbed from the samples.
A time-of-flight mass spectrometer,  mounted opposite to the hemispherical analyser, has been used to monitor the thermal stability of the sample at regular intervals during the measurements.
The experimental chamber is maintained at a background pressure of low 10$^{-8}$ mbar, reaching a few 10$^{-7}$ mbar in operating conditions. The target molecules, which are in the form of a powder at standard ambient temperature and pressure, are all commercially available. The c-TrpTrp and c-TrpTyr species were purchased from BACHEM while the c-GlyPhe molecule from Sigma-Aldrich. All the species have purity $\geq 98\%$ and have been used without further purification. In all cases, an amount of sample of about 30 mg was inserted in a crucible under vacuum and sublimated at a temperature of 95, 155 and 147$^\circ$C for c-GlyPhe, c-TrpTyr and c-TrpTrp, respectively.
The photoelectron spectra have been obtained by adding several scans over the binding energy region 7-30 eV with an energy step of 120 meV and acquisition time of 4 sec/point.

\section{Results}

\begin{figure}[h]
\centering
  \includegraphics[width=0.5\textwidth]{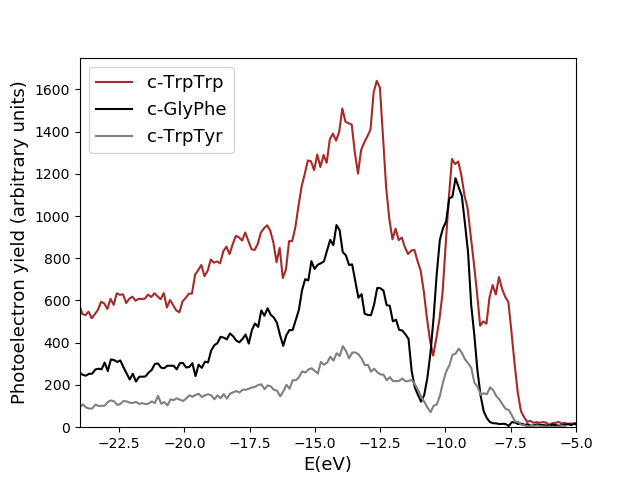}
  \caption{Experimental photoemission spectra of the c-GlyPhe, c-TrpTyr, and c-TrpTrp dipeptides. The feature at -12.6 eV in the c-TrpTrp spectrum is due to residual water in the sample.}
  \label{fig:3expPES}
\end{figure}

The experimental photoemission spectra of the c-GlyPhe, c-TrpTyr, and c-TrpTrp are shown in Figure \ref{fig:3expPES}. The structure of the three cyclo-dipeptides is made by the central DKP ring and different side chains depending on the constituent amino acids, which contain the phenyl, phenol and indole chromophore groups in Phe, Tyr and Trp respectively. Thus the three spectra share some general characteristics, namely a first feature in the 8 - 10 eV energy region, followed after a gap of a couple of eV by broader features which display  a maximum intensity at around 14 eV.
In the next sub-sections the {\it ab initio} simulations for each cyclo-dipeptide will be discussed in detail.

All amino acids except glycine (Gly) have a chiral center: each chiral molecule can exist in two chemically indistinguishable forms, enantiomers, labeled ``S'' and ``R'', which can only be identified through their different interaction with circularly polarized electromagnetic radiation, yielding opposite circular dichroism spectra. 
Out of the possible enantiomers or diastereoisomers resulting from the combinations of the ``S'' or ``R'' forms of the two constituent amino acids, we have performed our calculations - for each of the three investigated dipeptides (c-GlyPhe, c-TrpTyr, c-TrpTrp) - on the form on which experiments were performed, {\it i.e.} the ``S'' enantiomer for c-GlyPhe, and the ``S,S'' diastereoisomer for both c-TrpTyr and c-TrpTrp.

\subsection{ Cyclo(GlyPhe)}

\begin{figure}[h]
\centering
  \includegraphics[width=0.5\textwidth]{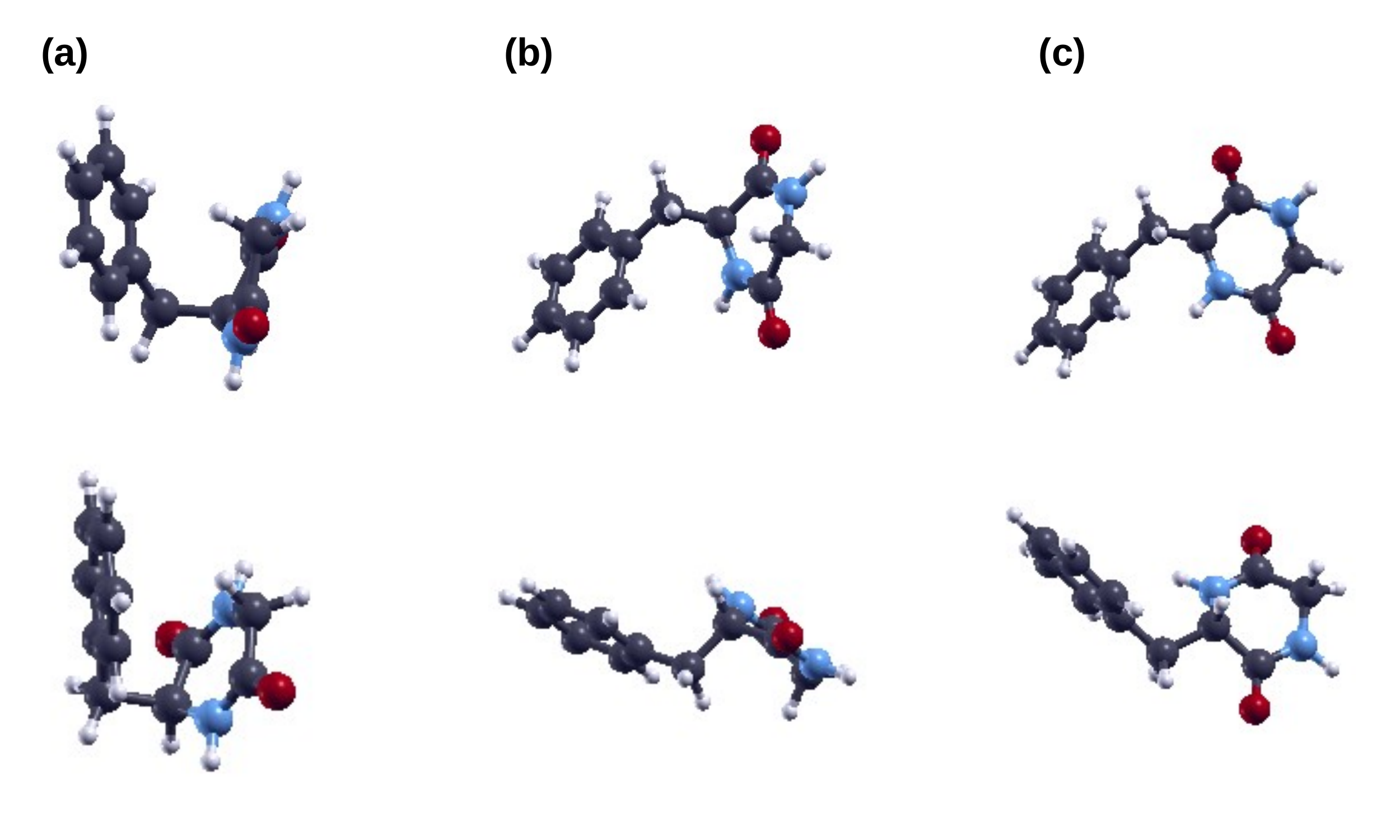}
  \caption{Optimized geometries of the three lowest energy conformers of the c-GlyPhe dipeptide, {\it i.e.} conformer 1 (a), conformer 2 (b) and conformer 3 (c), each shown in two different orientations. Carbon atoms are depicted in black, oxygen atoms in red, nitrogen atoms in light blue and hydrogen atoms in white.}
  \label{fig:geomGP}
\end{figure}

\begin{table}[h]
\small
 \caption{\ Tight-binding energies and populations, and ORCA B3LYP energies of the three lowest energy conformers of GlyPhe}
 \label{tab:E_TB_B3LYP_GlyPhe_all}
 \begin{tabular*}{0.48\textwidth}{@{\extracolsep{\fill}}llll}
  \hline
   &  TB $\Delta E$  & TB pop. &  ORCA $\Delta E$ (mHa) \\
   &  (mHa)          &  (\%)   &   B3LYP+vdW            \\
  \hline
  conf1  &  0.00   &  84.3 &  0.00    \\ 
  conf2  &  2.40  &  6.6  &  3.47   \\ 
  conf3  &  2.87  &  4.0  &  3.92  \\
    \hline
  \end{tabular*}
\end{table}

In Figure \ref{fig:geomGP} we show the three lowest energy conformers of the   cyclo(GlyPhe) peptide as obtained by the tight-binding conformational search using the CREST code. According to this method, the first conformer is much more populated than the following two conformers (second column of Table \ref{tab:E_TB_B3LYP_GlyPhe_all}).
Subsequent geometry relaxation within DFT B3LYP, both using the ORCA localized basis all-electron code (third column of Table \ref{tab:E_TB_B3LYP_GlyPhe_all}) and the QE plane-wave pseudopotential code, confirms the energy ordering of these conformers, and it does not substantially alter their geometry.

We use cyclo(GlyPhe), which is the smallest of the three peptides under study,
to compare the electronic properties with different levels of approximation for the exchange-correlation (xc) energy: HF, B3LYP and M062X, with the EOM-CCSD method, and with the GW approximation.
This was done in order to investigate the sensitivity of these properties to the chosen method and xc functional, and to find the method and/or functional which yields the best trade off between agreement with experimental data and computational cost. In particular, we focused on the comparison between measured and theoretically estimated - with different approaches - PES.

\begin{figure}[t!]
\centering
  \includegraphics[width=0.5\textwidth]{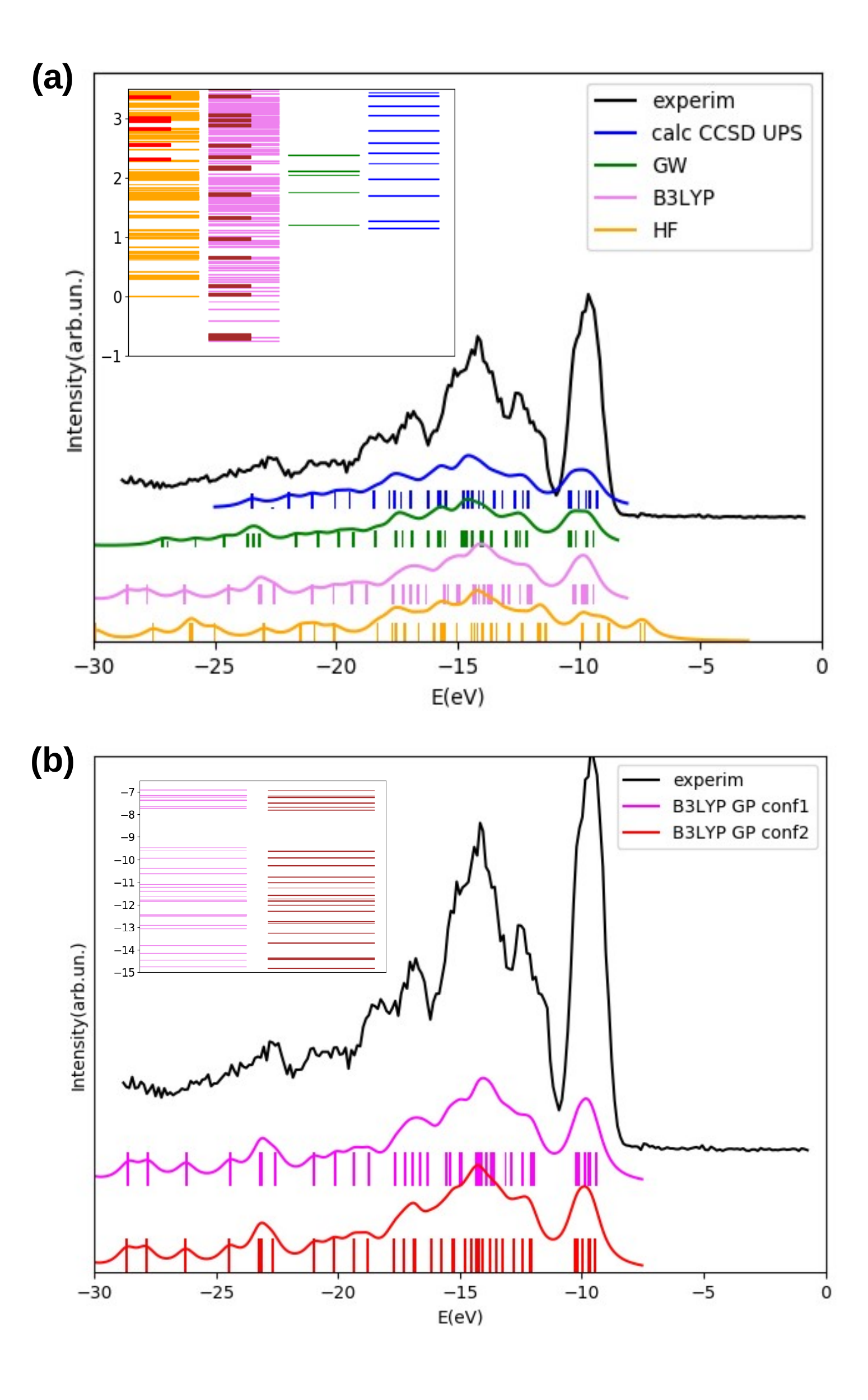}
  \caption{Panel (a): GlyPhe conformer 1. Hartree-Fock (yellow) and B3LYP (magenta) DOS for the occupied electronic levels, compared with the GW (green), and CCSD (blue) PE spectral function and with experimental PE spectrum (black). Inset: QE Hartree-Fock (yellow), ORCA Hartree-Fock (red), QE B3LYP (magenta), ORCA B3LYP (brown) unoccupied electronic energy levels compared with GW (green) and CCSD (blue) inverse PE energies. GW corrections have been computed for the five bound QE B3LYP empty levels only. The vacuum level is set at 0 energy.  
In the HF, B3LYP and GW cases occupied levels are shifted by +2 eV, -2.5 eV, -1.47 eV respectively, with empty levels unshifted.
Panel (b): DFT B3LYP DOS of GlyPhe conformer 1 (magenta) and conformer 2 (dark red), both shifted by -2.5 eV, compared to experimental photoemission spectrum (black); inset: DFT B3LYP occupied electronic energy levels of GlyPhe conformer 1 (magenta) and conformer 2 (dark red): the vacuum level is set at 0 energy.
The vertical axis in the two insets (as the horizontal one in the two large panels) reports energies in eV.
Curves for calculated DOS and PE spectral functions are obtained from the corresponding electronic energy levels by applying a Lorentzian broadening of 0.5 eV.}
  \label{fig:levGP}
\end{figure}

We inspect the electronic densities of states (DOS) for the GlyPhe conformer 1 (see Fig.~\ref{fig:levGP}, panel (a)) resulting from DFT B3LYP (magenta curve in the Figure) and HF (yellow curve) occupied electronic levels, and we compare the results with those obtained by calculating ionization energies using the EOM-CCSD method (blue) and the GW scheme (green curve). The latter are used as theoretical reference, and compared with the gas-phase experimental photoemission spectrum (black) of GlyPhe. In the EOM-CCSD DOS reported here each energy root has been weighed with its percentage contribution from single transitions, while in the GW simulation the quasi--particle poles $\epsilon_n$ are weighed by the renormalization factor $Z_n$.

We can observe that the B3LYP calculation provides a better description of the energy distribution of occupied electronic states with respect to the HF one, provided that a rigid shift of -2.5 eV is applied \footnote[3]{This is an extra shift on top of the vacuum level correction which we compute within the Makov-Payne and Martyna-Tuckerman (MT) schemes. For the box size used both schemes give $\approx 0.25$~eV} to the B3LYP levels in order to align the B3LYP HOMO level with the EOM-CCSD at $-9.25$~eV and thus to the experimental IE. We also report that the B3LYP IE obtained via total energy difference is ${E_0^{N-1}-E^N_0}=~8.57 $~eV, while the corresponding value obtained using the meta-GGA global hybrid M062X functional is 9.25 ~eV, thus matching the CCSD value. 
HF tends to overbind all main features of the PE spectrum by $\approx -2$~eV. This is why a rigid shift in the opposite direction is applied to the plot of the occupied HF levels. However, apart from this rigid shift, the agreement between the distribution of HF levels and the PE spectrum is not good, and HF tends to overestimate the energy splitting of the levels close to the HOMO of $\approx +2$~eV.
As a consequence, the unshifted $\epsilon^{HF}_{HOMO}=9.29$~eV is very close to the CCSD result, but this is the result of two errors with opposite sign. For better clarity the unshifted HOMO/IE are reported in Tab.~\ref{tab:IE_vs_EA} and later discussed.

The EOM-CCSD method, taking into account dynamical electronic correlation, is able to yield a PE spectral function (blue curve) in good agreement with the experimental PE spectrum, despite starting from the Hartree-Fock calculation which by itself badly describes the energy distribution of electronic levels. The EOM-CCSD PE energies and the B3LYP DOS are in a surprising and satisfactory agreement with each other in the analyzed energy range, although the CCSD scheme is numerically much more demanding.
The GW corrections (green curve) on top of the B3LYP energies lead to a final result which is even closer to the the EOM--CCSD one, in particular in the width of the first DOS feature (the one at $\approx$ -10 eV) and in the detailed shape of the various peaks/features in the -11 eV to -19 eV range. 
GW also partially cancels the underestimation of the HOMO binding energy. A residual underestimation of -1.47~eV remains. The GW levels in Fig.~\ref{fig:levGP} are then shifted by such value. Finally GW and EOM-CCSD also agree for the renormalization of the PE peaks intensity, determined by the percentage contribution from single transitions in EOM-CCSD and by the renormalization factor $Z_n$ in GW.
In the EOM--CCSD approach, contributions from double transitions are negligible up to $\approx -19$~eV (an energy close to the double ionization threshold \cite{Eland_Photoion}), and the mixing between single and double transitions is very small in the whole analyzed energy range, ({\it i.e.}, up to $\approx -25$~eV). With the exception of two energy roots, with percentage contributions from single transitions of 80$\%$, $\approx$10$\%$ respectively, the contribution from single transitions to a given energy root is either larger than 90\%, or  below 0.5$\%$. Similarly the renormalization factors $Z_{n}$ of the GW calculation are all above 0.75 up to $\approx -23$~eV.

In panel (b) of Figure \ref{fig:levGP} we compare DFT B3LYP occupied energy levels and DOS of the two lowest energy conformers of the GlyPhe peptide. Although the geometries of these two conformers are quite different (Fig. \ref{fig:geomGP}), the energy positions of their electronic levels (inset of Fig.~\ref{fig:levGP} (b)) display minor mutual differences only. Once a Lorentzian broadening is applied to these electronic levels in order to obtain the densities of states of the two conformers, these latter appear hardly distinguishable from each other, being both in good agreement with the features in the PE spectrum. 

\begin{table}[h]
\small
 \caption{\ Ionization energy (IE) and the opposite of the electron affinity (-EA) of GlyPhe conformer 1 with different methods. Values are in eV. The experimental value (column 6) is a vertical IE and it corresponds to the maximum of the first feature in the photoelectron spectrum. Also the IE and EA values calculated via DFT, with either the B3LYP (column 2) or the M062X (column 3) exchange-correlation functional, both with the def2-QZVPP basis set and the D3 vdW correction, are vertical values.
 In the Supporting Information$\dag$ we report further IE and EA values for the three investigated cyclo-dipeptides, calculated via total energy differences, including a comparison between vertical and adiabatic values}
 \label{tab:IE_vs_EA}
 \begin{tabular*}{0.48\textwidth}{@{\extracolsep{\fill}}lllllll}
  \hline
        & HF     & B3LYP &  M062X & GW     & CCSD  & exp   \\
        & eigen. & Etot  &  Etot     & eigen. & roots &       \\
  \hline
  IE    &  9.29    & 8.57  &  9.25  &  7.94  &  9.25   &  9.54±0.02 \\ 
 -EA    &   2.30    &  0.81  & 0.91  &  1.21    &  1.15     &    \\ 
   gap  &   11.59    &  9.38 &  10.16  &  9.15  &   10.4    &    \\
    \hline
  \end{tabular*}
\end{table}

We now focus on empty electronic levels.
The inset of Figure~\ref{fig:levGP}, panel (a), shows the position of unoccupied electronic energy levels of GlyPhe conformer 1 obtained within Hartree-Fock (HF) and B3LYP, compared with the inverse PE energies obtained within the GW method on top of B3LYP and within the EOM-CCSD method. For the GW method we only consider the correction to the QE B3LYP bound empty states. No energy shift is applied to the empty states besides the vacuum level correction computed within HF (0.18~eV on the cubic cell) and B3LYP (0.18~eV on the cubic cell; 0.25~eV on the FCC cell). This is somehow equivalent to assume that the shifts applied to the occupied states correspond to an opening of the band gap, {\it i.e.} the energy difference between the IE and minus the EA.
EOM-CCSD gives a negative EA, {\it i.e.} the first root has positive energy of 1.15~eV (fifth column in Table~\ref{tab:IE_vs_EA}). In contrast B3LYP gives $\epsilon^{KS}_{n}< 0$ for $n=LUMO$ and 5 bound empty states in total. The B3LYP EA obtained via total energy difference is instead negative ${E_0^{N}-E^{N+1}_0}=-0.81$~eV. At the HF level the first unoccupied state localized on the molecule, obtained from the simulation with localized basis set, is at $\epsilon^{HF}_{LUMO}\approx 2.3$~eV. We refer to such state as a resonance; in the ESI$\dag$ we show the spatial localization of this state. In the figure inset, HF resonances from ORCA (with localized basis set) are highlighted in red in the dense list of states obtained with QE using the plane-waves basis set, which also captures vacuum states. In a similar way, B3LYP empty levels obtained with ORCA are highlighted in brown in the list of B3LYP empty states obtained with QE. In the B3LYP case we obtain respectively five (two) bound empty states with QE (ORCA), but without a significant variation in the energy of the first unoccupied state (-0.75 eV with QE B3LYP, -0.70 eV with ORCA B3LYP).
The GW correction to the bound B3LYP empty levels also gives a negative EA with  $\epsilon^{GW}_{LUMO}= 1.21$~eV, quite close to the EOM-CCSD result. GW corrections to the delocalized vacuum states are not considered.
Similarly to the occupied levels case, in the EOM-CCSD approach, the contribution from single transitions is larger than 90$\%$ for all the inverse PE energy roots, and in GW the renormalization factors $Z_{n}$ are larger than 0.9 for all the computed empty levels.

We underline that in the positive energy range it is not possible to compare the energy levels distribution between the different approaches due to the presence of delocalized states from the continuum. Continuum states are captured when using the plane--waves basis set. On the other hand the approach based on localized basis set only captures resonances, although missing their delocalized contribution (resonances have positive energy, as a consequence they must be in part delocalized) and thus possibly introducing an error in the resulting energy. A good description of resonances using a plane--waves basis set would require very large boxes or possibly some lifetime to accelerate convergence. Still the comparison between results with the two basis sets is interesting to be investigated also with the finite boxes we use (see Supplementary Information\dag).

The overall picture is that GlyPhe has a negative EA of $1.15$~eV, a IE of $9.25$~eV and a gap, or difference between IE and EA, of $10.4$~eV. In Table \ref{tab:IE_vs_EA} we report the IE and the EA from the different approaches, assuming that HF and B3LYP eigenvalues for the HOMO and the LUMO are a zero order approximation. B3LYP eigenvalues, not shown in the Table, underestimate both the IE (6.91 eV) and EA (0.75 eV), and as a consequence the gap, while HF does the opposite.\cite{Mori-Sanchez_PRL2008} B3LYP and M062X total energies and GW eigenvalues still underestimate the gap and the IE and EA, but the error is much smaller. In particular, the meta-GGA global hybrid M062X functional yields vertical IE and EA values in better agreement than B3LYP with those obtained with the CCSD method, and therefore a IE value which is closer to the experimental one (also vertical). The different results are mostly affected by how the electron exchange or Fock term enter in the description. HF has the full exchange term which opens too much the electronic gap, while B3LYP has only a fraction of the exchange which is not enough. Both GW and CCSD add extra terms which lead to a screening of the exchange (in GW {\it via} the RPA screening and in CCSD {\it via} the inclusion of terms which mimic the IP screening) leading to the correct inclusion of this effect. The dynamical corrections included both within GW and CCSD do not play a significant role since all poles have dominant single character (/are well described within the QP approximation). However CCSD also includes extra terms, neglected in GW, that give a sizable effect. These extra terms are known to be important in small molecules, but their relevance usually decreases while increasing the system size (up to the the limit of extended systems where GW gives a very good physical description). Finally, while the use of screened exchange and extra terms is important for a correct electronic gap, the energy levels distribution is much less sensitive as already discussed.

\begin{figure}[h]
\centering
  \includegraphics[width=0.5\textwidth]{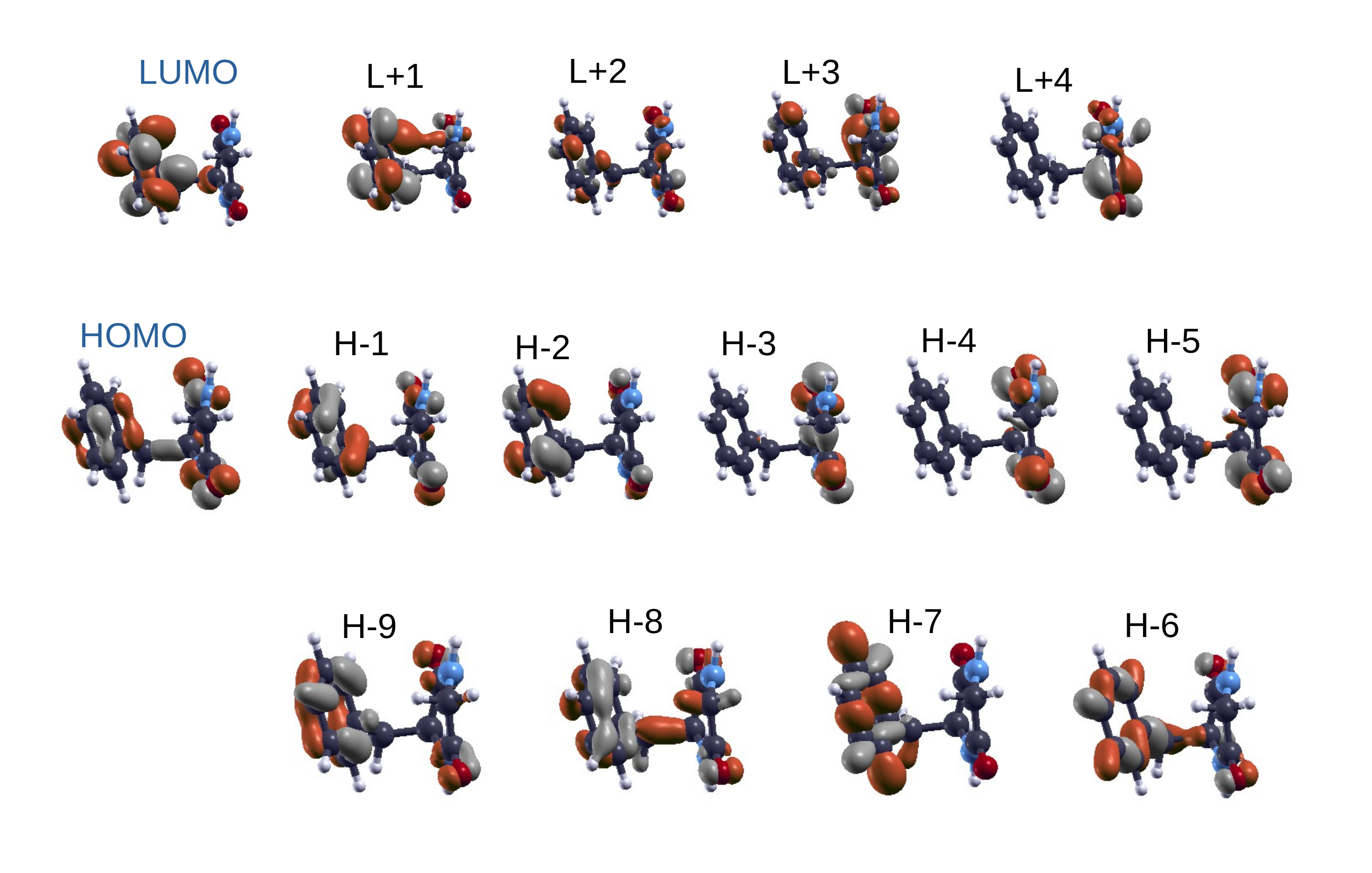}
  \caption{DFT B3LYP orbitals of GlyPhe conformer 1.}
  \label{fig:WF_B3LYP_GP1}
\end{figure}

In order to shed light on similarities between the EOM-CCSD on top of HF methods and the GW on top of B3LYP scheme, we also analyzed the spatial localization of some B3LYP occupied and empty electronic states in the energy region of the frontier orbitals, encompassing the gap, as shown in Figure \ref{fig:WF_B3LYP_GP1} for the GlyPhe conformer 1.
Each of the EOM-CCSD ionization energies can be interpreted as a linear combination of ionizations of the subsiding Hartree-Fock levels. We thus compare the spatial localization of the HOMO B3LYP bound electronic states with that of the weighed superposition of the HF states mostly contributing to the EOM-CCSD ionization energies.

\begin{figure}[h]
\centering
  \includegraphics[width=0.5\textwidth]{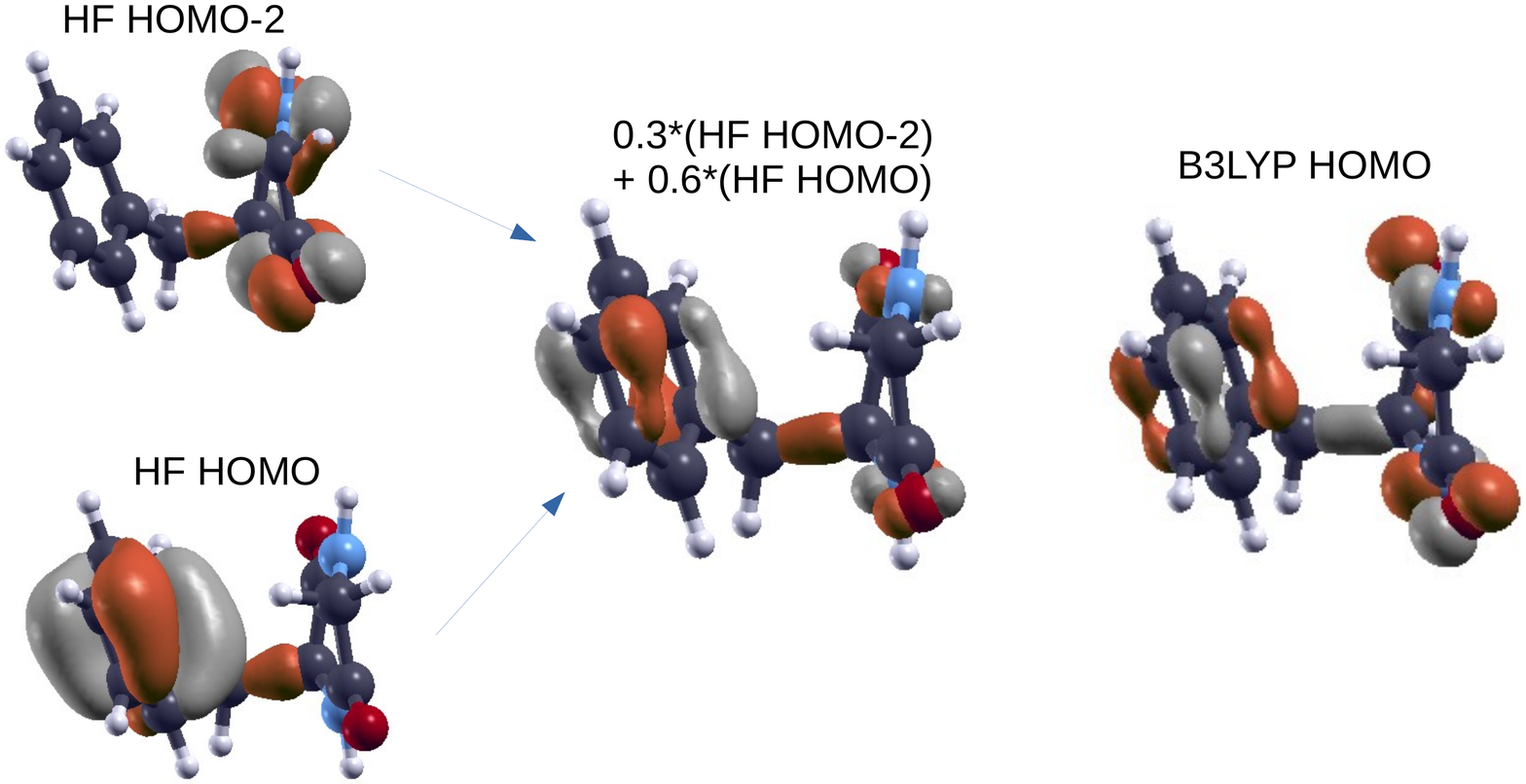}
  \caption{GlyPhe conformer 1: Hartree-Fock orbitals (left) contributing to the CCSD ionization energy (center), and the B3LYP HOMO (right).}
  \label{fig:WF_CCSD-B3LYP}
\end{figure}

As shown in Figure \ref{fig:WF_CCSD-B3LYP}, the agreement is very good with the first EOM-CCSD ionization energy mixing HOMO and HOMO-2 HF orbitals to lead to a final wave-function which is very similar to the B3LYP HOMO. We did the same analysis considering also the following 2 EOM-CCSD PE energies which we label here IE-1 and IE-2 (Supplementary Information).$\dag$ IE-1 mixes HOMO, HOMO-1 and HOMO-2 HF orbitals leading to a wave--function very close to the B3LYP HOMO-2 state.
IE-2 mixes HOMO-5, HOMO-2, HOMO-1 and HOMO HF orbitals, leading to a wave--function similar (but with some differences) to the B3LYP HOMO-5 state. 
In conclusion, from the analysis of both electronic densities of states and spatial localization of molecular orbitals, we can thus conclude that DFT B3LYP calculations provide a reasonably good description of the electronic properties, if compared to the EOM-CCSD approach, but at a much lower computational cost. 
This is why we take the liberty of use the wording ``molecular orbitals'' also for the B3LYP ones.

\subsection{Cyclo(TrpTyr)}
We now turn our attention to the TrpTyr molecule. As opposed to GlyPhe, TrpTyr has two chiral centers, one on each of the two constituent amino acids. Therefore, in addition to chemically indistinguishable enantiomers of the TrpTyr dipeptide, {\it i.e.} pairs of molecule isomers each of which is the mirror image of the other (such as SS vs. RR), we can also have pairs of chemically distinguishable diastereoisomers, such as SS vs. SR, where {\it e.g.} ``SR'' indicates a TrpTyr molecule with Trp in its ``S'' enantiomer and Tyr in its ``R'' one. For our calculations we chose the ``SS'' diastereoisomer, the one used in the photoemission measurements.

\begin{figure}[h]
\centering
  \includegraphics[width=0.5\textwidth]{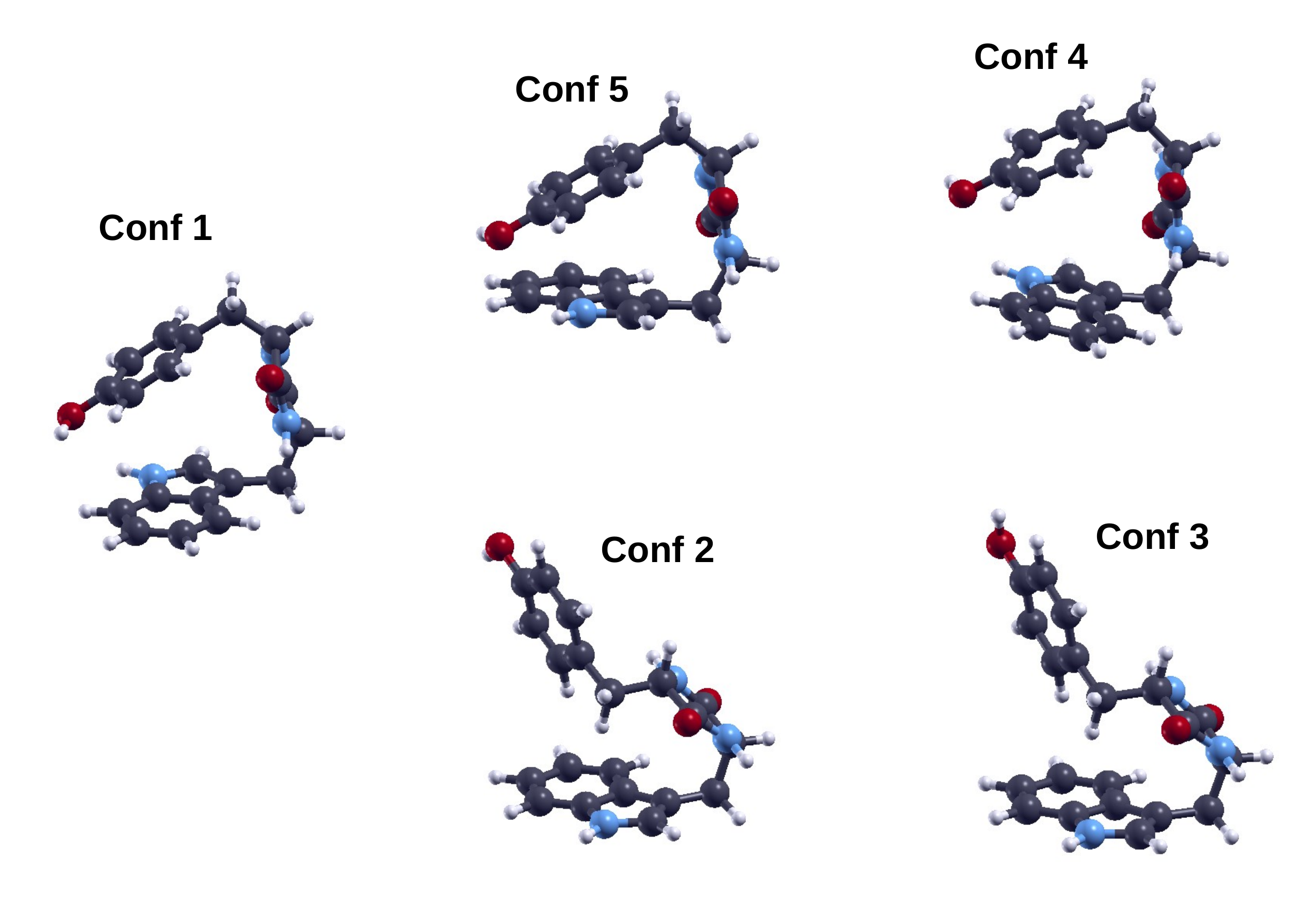}
  \caption{Geometries of the five lowest energy conformers of the cyclo(TrpTyr) peptide. Color codes as in Figure \ref{fig:geomGP}.}
  \label{fig:geomTrpTyr}
\end{figure}

In Figure \ref{fig:geomTrpTyr} we show the five lowest energy conformers of the ``S,S'' diastereoisomer of the cyclo(TrpTyr) peptide as obtained by a tight-binding conformational search using the CREST code. Conformers are labeled (1, 2, ..) according to their tight-binding energy ordering (column 1 of Table \ref{tab:E_TB_B3LYP_TrpTyr_all}).
For TrpTyr the energy ordering of the conformers obtained through the tight-binding run changes upon performing a DFT B3LYP geometry relaxation, with either the ORCA or the QE code, although their geometries remain substantially unaltered. 
In particular, the energy ordering obtained after performing ORCA B3LYP relaxations (column 5 of Table \ref{tab:E_TB_B3LYP_TrpTyr_all}) is the following (using conformer labels based on their tight-binding energy ordering): 1, 5, 4, 2, 3. Thus conformer 1 remains the lowest energy geometry, while conformers 5 and 4, in which (as in conformer 1) the side-chain rings of the tyrosine and tryptophan amino acids are closer to each other, become energetically favored with respect to conformers 2 and 3. Based on these results, we decided to focus on conformers 1, 5 and 4 for further calculations on electronic properties. 

\begin{table}[h]
\small
 \caption{\ Tight-binding energy differences and populations, and ORCA B3LYP energies of the five lowest energy conformers of TrpTyr }
 \label{tab:E_TB_B3LYP_TrpTyr_all}
 \begin{tabular*}{0.48\textwidth}{@{\extracolsep{\fill}}llllll}
  \hline
        &  TB $\Delta E$  & TB pop. &  \multicolumn{3}{c|}{ORCA $\Delta E$ (mHa)} \\
         &  (mHa)          &  (\%) &  B3LYP & \,vdW\, & \,sum\, \\
  \hline
  conf1  &  0.00  & 30.4   & 0.00 & 0.00 &   0.00  \\ 
  conf2  &  0.38  & 20.3   & 0.32 & 2.52 &  2.84   \\ 
  conf3  &  0.57  & 16.6   & 0.59 & 2.57 &  3.16   \\
  conf4  &  1.14  &  0.09  & 0.26 & 1.79 &  2.06   \\
  conf5  &  1.49  &  0.06  & 1.49 & -0.49&  1.00   \\
    \hline
  \end{tabular*}
\end{table}

Intramolecular weak interactions such as the van der Waals forces can have an important role in molecule stability, and the chosen method for their computational treatment can alter the energy ordering of molecule conformers. $\pi$ - $\pi$ and CH - $\pi$ interactions can contribute to stabilizing specific conformers, in molecules containing aromatic rings. Indeed, out of the five low-energy geometries obtained through the tight-binding conformational search, conformers 1, 5 and 4 appear to be stabilized to some extent by $\pi$ - $\pi$ interactions between the phenol ring of tyrosine and the indole ring of tryptophan, while the geometries of conformers 2 and 3 suggest a possible CH - $\pi$ interaction between the C$_{\beta}$ group of tyrosine and the indole ring of tryptophan.
Interestingly, the dispersive vdW contribution (column 4) to the computed total energy of c-TrpTyr has an important role in determining the DFT B3LYP energy ordering of its conformers (see column 5 of Table~\ref{tab:E_TB_B3LYP_TrpTyr_all}). We recall that a similar vdW correction is present in the TB simulation.
More in detail: the ORCA B3LYP calculation (third column in the Table) tends to energetically favor conformer 4 with respect to the energy ordering obtained from the tight-binding conformational search (first column) as long as we do not consider the vdW contribution to the DFT total energy. The vdW contribution to the DFT energy (fourth column) is negative for conformer 5 only - and zero or positive for the other conformers - thus strongly favoring conformer 5. As a final result, the energy ordering of conformers according to the total B3LYP energy including the vdW contribution is, as already mentioned, 1, 5, 4, 2, 3. The same final energy ordering is also obtained using the D4 vdW dispersion correction \cite{vdW_D4_JCP2017,vdW_D4_JCP2019} in the ORCA B3LYP calculation, instead of the D3 discussed so far: the only remarkable difference is that the ORCA calculation with the D4 dispersion correction yields a positive vdW contribution for conformer 5; this is not sufficient to change the energy ordering of the conformers, but it brings the energy of conformer 5 very close to that of conformer 4 (differing by ~1 meV only).

\begin{figure}[h]
\centering
  \includegraphics[width=0.5\textwidth]{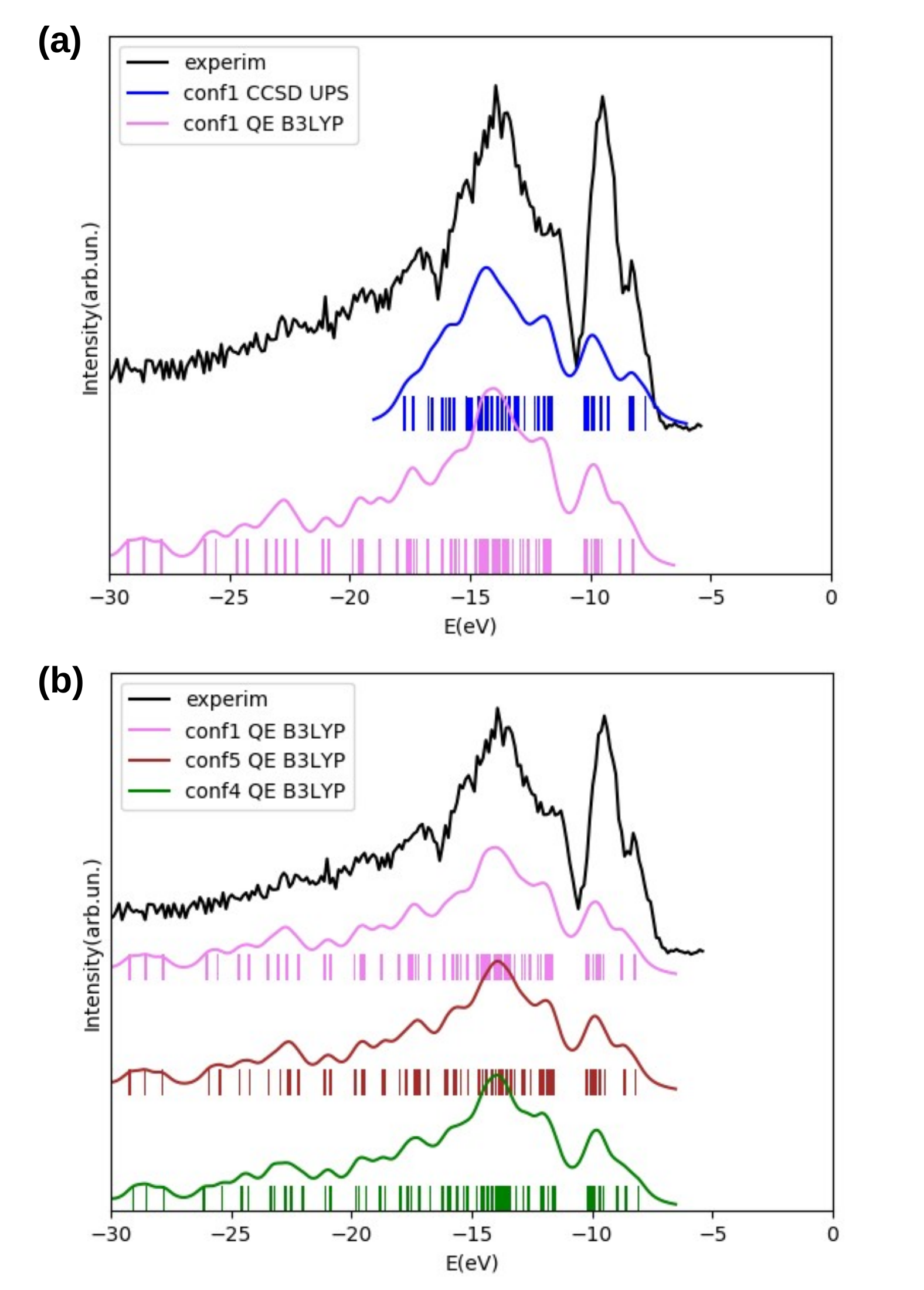}
  \caption{Panel (a): DOS of TrpTyr conformer 1 obtained with the CCSD method (blue), and with DFT B3LYP (magenta), compared to experimental photoemission spectrum of TrpTyr (black): the B3LYP DOS is shifted by by -2.5 eV. Panel (b): DFT B3LYP DOS of TrpTyr conformer 1 (magenta), conformer 5 (dark red) and conformer 4 (green), all shifted by -2.5 eV, compared to experimental photoemission spectrum (black). The horizontal axes of both panels report energies in eV. Curves for calculated DOS and PE spectral functions are obtained from the corresponding electronic energy levels by applying a Lorentzian broadening of 0.5 eV.}
  \label{fig:levTrpTyr_conf}
\end{figure}

In Figure \ref{fig:levTrpTyr_conf}, panel (a), we show the electronic densities of occupied states for TrpTyr conformer 1 ({\it i.e.} the most populated one according to both tight-binding conformational search and B3LYP DFT) obtained either from DFT B3LYP (magenta curve), or from a valence photoemission spectrum (blue) obtained through a EOM-CCSD calculation. Also for this cyclo-dipeptide, as already observed for GlyPhe, the DFT B3LYP and EOM-CCSD densities of states are in good agreement with each other and with the experimental photoemission spectrum (black curve), once a -2.5~eV energy shift is applied to the B3LYP DOS. Also here (as in the GlyPhe case) the EOM-CCSD energy roots have been weighed with their percentage contribution from single transitions. For TrpTyr this percentage is higher than 90 in all the analyzed energy range, which is indeed shorter than the one considered for GlyPhe (since considering the same number of energy roots for a larger molecule). 

As for the sensitivity of the DFT B3LYP densities of states to molecule conformation, panel (b) of the same figure shows the DFT B3LYP DOS for the three lowest energy conformers of TrpTyr according to DFT B3LYP, {\it i.e.} conformers 1, 5 and 4 according to tight-binding labeling. Also for TrpTyr we can only observe minor differences among the densities of states of different conformers, such as a shoulder at $\approx$ 8 eV ionization energy, which is most pronounced in conformer 5, less in conformer 1, and almost absent in conformer 4. As such, also for TrpTyr the densities of states of all the three analyzed conformers are in good agreement with the experimentally measured photoemission spectrum of the molecule (black curve in the Figure).

\begin{figure}[h]
\centering
 \includegraphics[width=0.5\textwidth]{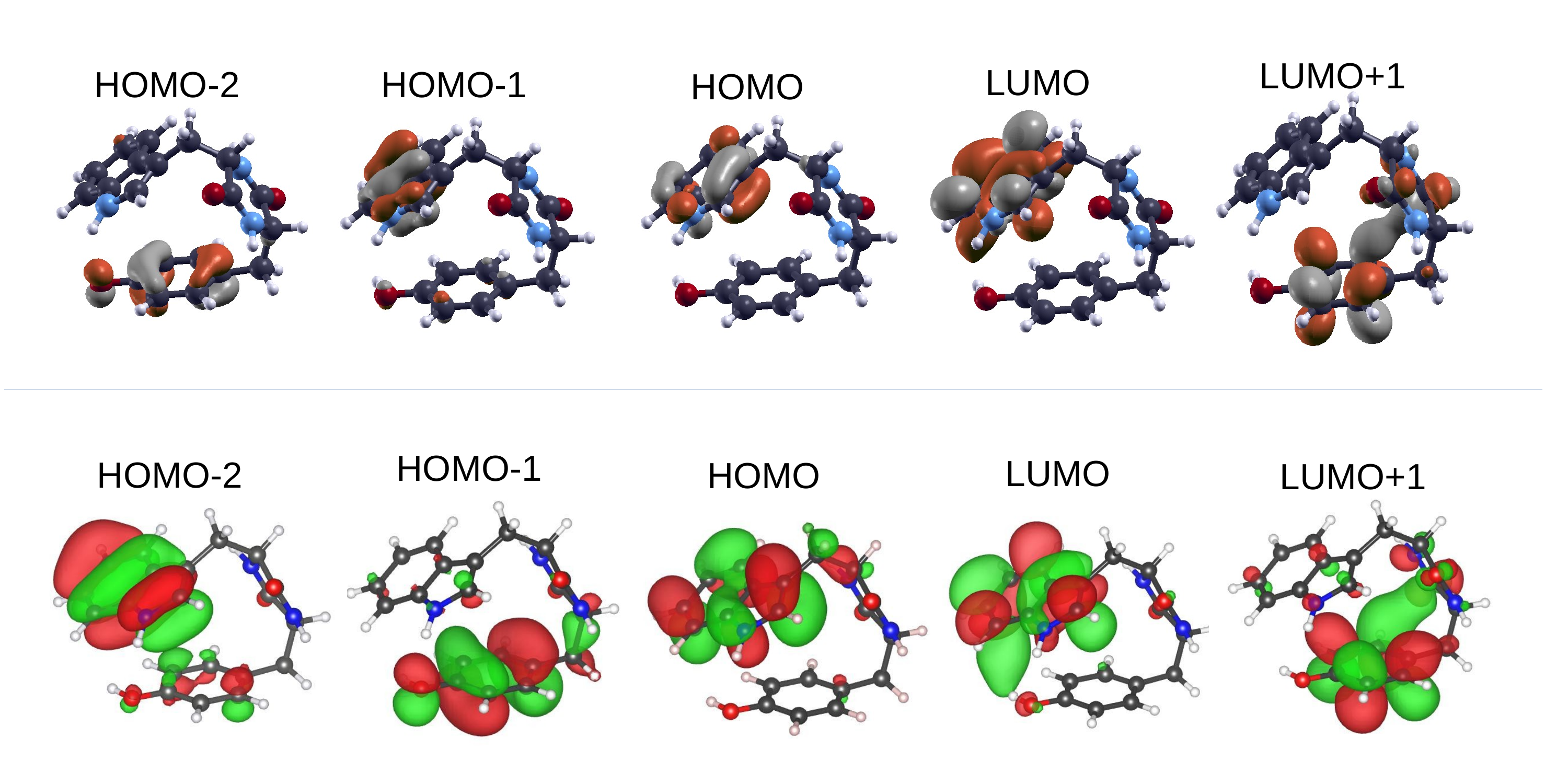}
  \caption{QE (upper panel) and ORCA (lower panel) B3LYP orbitals of TrpTyr conformer 1.}
  \label{fig:WF_TrpTyr1_QE_orca}
\end{figure}

The spatial localization of electronic states of TrpTyr lowest energy conformer (conformer 1) is shown in Figure \ref{fig:WF_TrpTyr1_QE_orca}. Frontier orbitals only are considered, calculated within DFT B3LYP with either the plane-wave pseudopotential code QE or the localized basis all-electron code ORCA.
As expected, the two codes give almost identical results, apart from a switch between HOMO-2 and HOMO-1 which are very close in energy (Table \ref{tab:E_neargaplev_TrpTyr}).

From Table~\ref{tab:E_neargaplev_TrpTyr} we can also observe that the ORCA B3LYP calculation yields two bound empty states only (column 2); on the other hand, in the QE B3LYP calculation we find seven bound empty states, {\it i.e.} three further states in addition to the ones appearing in column 1 of the Table. As already observed for the c-GlyPhe dipeptide, in the ORCA B3LYP calculation with localized bases empty levels are generally less ``dense'' in energy with respect to those obtained from the plane-waves QE B3LYP calculation, while the energy distribution of occupied states is very similar with the two approaches. This trend can be rationalized by recalling that DFT calculations with plane-waves basis sets are able to also capture continuum(-like) delocalized empty electronic states, as opposed to localized basis codes, which only capture resonances, possibly introducing an error in their energies.

\begin{table}[h]
\small
 \caption{\ Energies of the highest occupied and of the lowest unoccupied QE B3LYP and ORCA B3LYP electronic levels for conformer 1 of TrpTyr. Occupied levels are shifted by -2.5 eV.}
 \label{tab:E_neargaplev_TrpTyr}
 \begin{tabular*}{0.48\textwidth}{@{\extracolsep{\fill}}lll}
  \hline
        &  QE (eV)  & ORCA (eV)    \\  
  \hline
   LUMO+3(*) &  -0.3325   &   +0.2208   \\
   LUMO+2(*) &  -0.4689   &   +0.0528   \\ 
   LUMO+1   &  -0.6307   &   -0.3518    \\ 
   LUMO     &  -0.7573   &   -0.5886  \\ 
   HOMO     &  -8.2152   &   -8.0903   \\ 
   HOMO-1   &  -8.7344   &   -8.5768   \\ 
   HOMO-2   &  -8.7602   &   -8.6369   \\ 
    \hline
  \end{tabular*}
\end{table}

\subsection{Cyclo(TrpTrp)}
The TrpTrp dipeptide has two chiral centers (one for each of the two constituent amino acids) and therefore, as already mentioned for TrpTyr, the four combinations of enantiomers of the single amino acids ``S,S'', ``R,R'', ``S,R'', ``R,S'' are possible. Being TrpTrp composed of two identical molecules (at a difference with the other two analyzed dipeptides), the ``S,R'' and ``R,S'' isomers may in principle be achiral due to symmetry reasons, however their symmetry and therefore chirality properties will in general depend on the conformation. In our calculations we considered the ``SS'' diastereoisomer, {\it i.e} the one used in the photoemission measurements.

\begin{figure}[t]
\centering
 \includegraphics[width=0.5\textwidth]{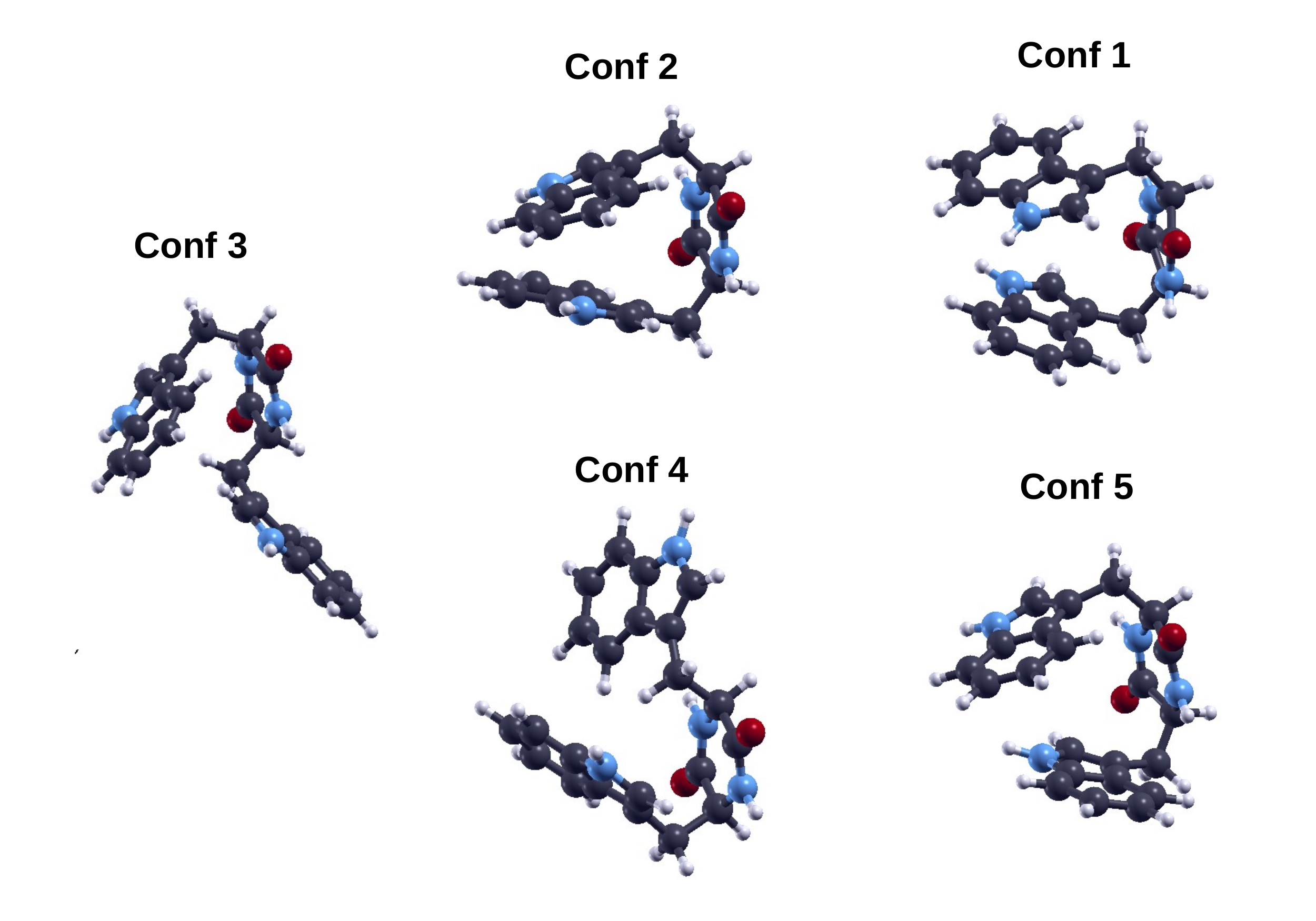}
  \caption{Geometries of the five lowest energy conformers of the cyclo(TrpTrp) peptide. Color codes as in Figure \ref{fig:geomGP}.}
  \label{fig:geom_TrpTrp}
\end{figure}

Also for the ``S,S'' form of TrpTrp, as already found for TrpTyr, the energy ordering of conformers as obtained through the initial TB conformational search is not maintained in QE DFT B3LYP (see columns 1 and 3 of Table \ref{tab:E_TB_B3LYP_TrpTrp_all}). In this case not even the lowest energy conformer is conserved: labeling conformers according to their TB energy order, within DFT B3LYP with the QE code the lowest energy geometry is conformer 3, followed by conformers 2, 1, 4 and 5 (their geometries are shown in Figure \ref{fig:geom_TrpTrp}). Also for this dipeptide, some of the five TB lowest energy conformers, {\it i.e.} conformers 2, 1 and 5, appear likely to be stabilized to some extent by $\pi$ - $\pi$ interactions between the aromatic rings of the two constituent amino acids - here two tryptophan indoles - and others (conformers 3 and 4) suggest CH - $\pi$ interactions between the C$_{\beta}$ group of one tryptophan and the indole ring of the other.

\begin{table}[h]
\small
 \caption{\ Tight-binding energies and populations, and QE B3LYP energies of the five lowest energy conformers of TrpTrp}
 \label{tab:E_TB_B3LYP_TrpTrp_all}
 \begin{tabular*}{0.48\textwidth}{@{\extracolsep{\fill}}llll}
  \hline
         &  TB $\Delta E$  & TB pop. & QE $\Delta E$ (mHa) \\
         &  (mHa)          &  (\%) &    B3LYP+vdW \\
  \hline
  conf1  &  0.00  & 35.1  &  0.00      \\ 
  conf2  &  0.52  & 20.3  & -0.15     \\ 
  conf3  &  1.21  & 19.5  & -0.41    \\
  conf4  &  1.51  & 14.3  &  1.74    \\
  conf5  &  2.53  &  4.8  &  5.00    \\
    \hline
  \end{tabular*}
\end{table}

A more detailed analysis (see ESI\dag), exploring the dependence of the DFT total energy ordering of the five lowest energy conformers from the tight binding conformational search on the computational details of the DFT calculation, such as cell size, localized bases (ORCA) vs. plane waves (QE), type of vdW treatment, exchange-correlation functional, pseudopotentials, has shown that the energy ordering among conformers 1, 2 and 3 can change in some cases, but these three conformers are always lower in energy than conformers 4 and 5. Moreover, the energy differences among conformers 1, 2 and 3 are in most cases within the value of $k_{B}T$ at room temperature, and lower than those between these three conformers and the other two (conformers 4 and 5). This supports our choice to analyze conformers 1, 2 and 3 as lowest energy / most populated conformers of this peptide. 

\begin{figure}[t!]
\centering
 \includegraphics[width=0.5\textwidth]{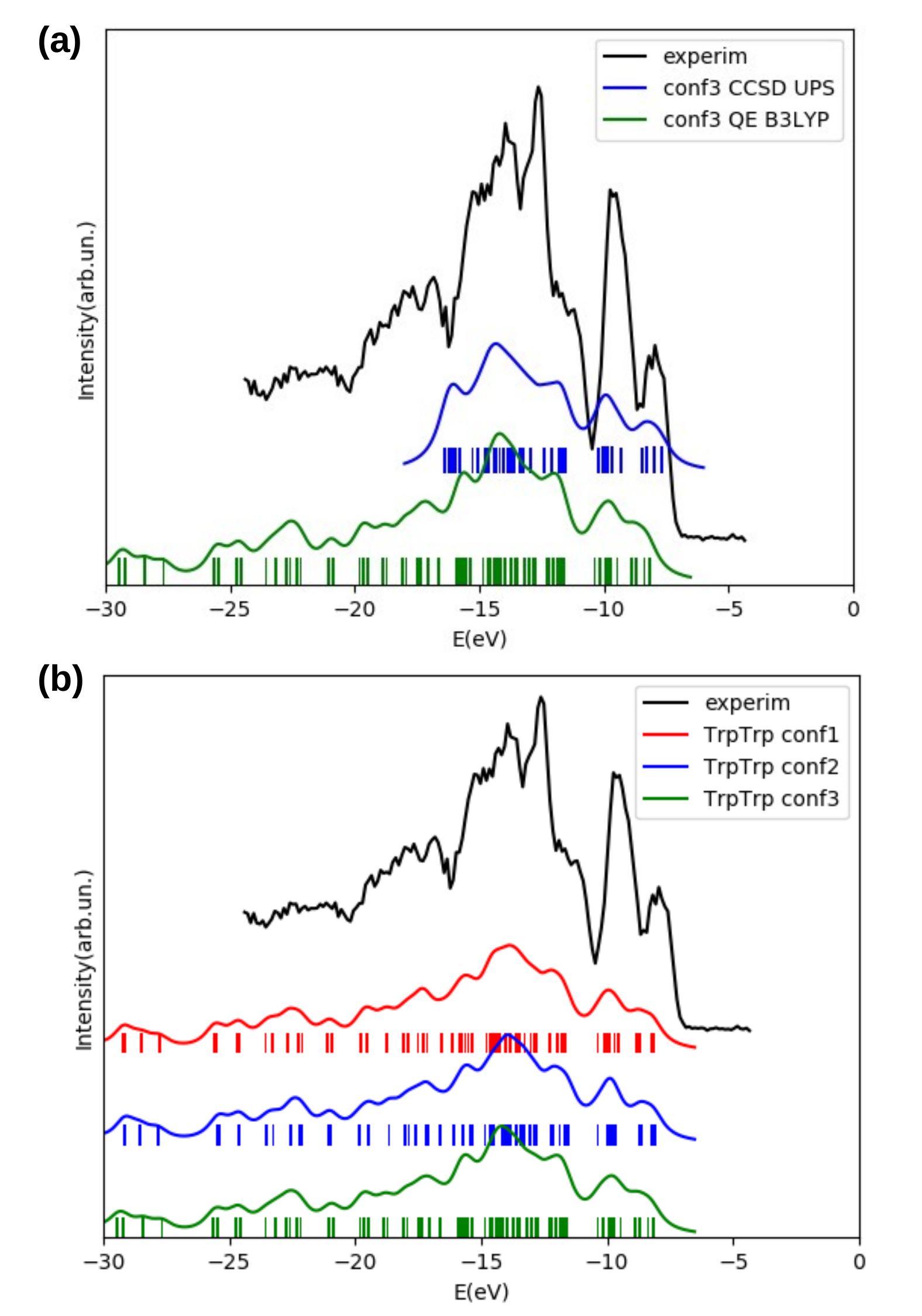}
  \caption{Panel (a): DOS of TrpTrp conformer 3 obtained with the EOM-CCSD method (blue), and with DFT B3LYP (green), compared to the experimental photoemission spectrum of TrpTrp (black): the B3LYP DOS is shifted by -2.5 eV. Panel (b): DFT B3LYP DOS of TrpTrp conformer 1 (red), conformer 2 (blue) and conformer 3 (green), all shifted by -2.5 eV, compared to experimental photoemission spectrum (black). The feature at -12.6 eV in the experimental spectrum is due to residual water in the sample. The horizontal axes of both panels report energies in eV. Curves for calculated DOS and PE spectral functions are obtained from the corresponding electronic energy levels by applying a Lorentzian broadening of 0.5 eV.}
  \label{fig:levTrpTrp_conf}
\end{figure}

The agreement between the electronic densities of states obtained either from DFT B3LYP energy levels (green curve in panel (a) of Figure~\ref{fig:levTrpTrp_conf}) or from a valence photoemission spectrum (blue curve) obtained through a EOM-CCSD calculation is quite good, and both curves (obtained for the lowest energy conformer according to DFT B3LYP) reproduce reasonably well the experimental photoemission spectrum (black) of TrpTrp in the analyzed energy range. The EOM-CCSD energy roots have been weighed with the percentage contribution from single transitions, which is also in this case (as in TrpTyr) higher than 90 for all the obtained ionization energies.

A comparison between DFT B3LYP densities of states of the three lowest energy conformers of the TrpTrp dipeptide (panel (b) of Figure~\ref{fig:levTrpTrp_conf}) shows a negligible conformational sensitivity of the DOS also for this molecule: the three DOS curves are almost indistinguishable, despite the rather different geometries of the analyzed conformers (Fig.~\ref{fig:geom_TrpTrp}). Also in this case, the calculated density of states alone would not be sufficient in order to draw any conclusions on the presence of the different possible conformers in the experimental sample.

\begin{figure}[h]
\centering
 \includegraphics[width=0.5\textwidth]{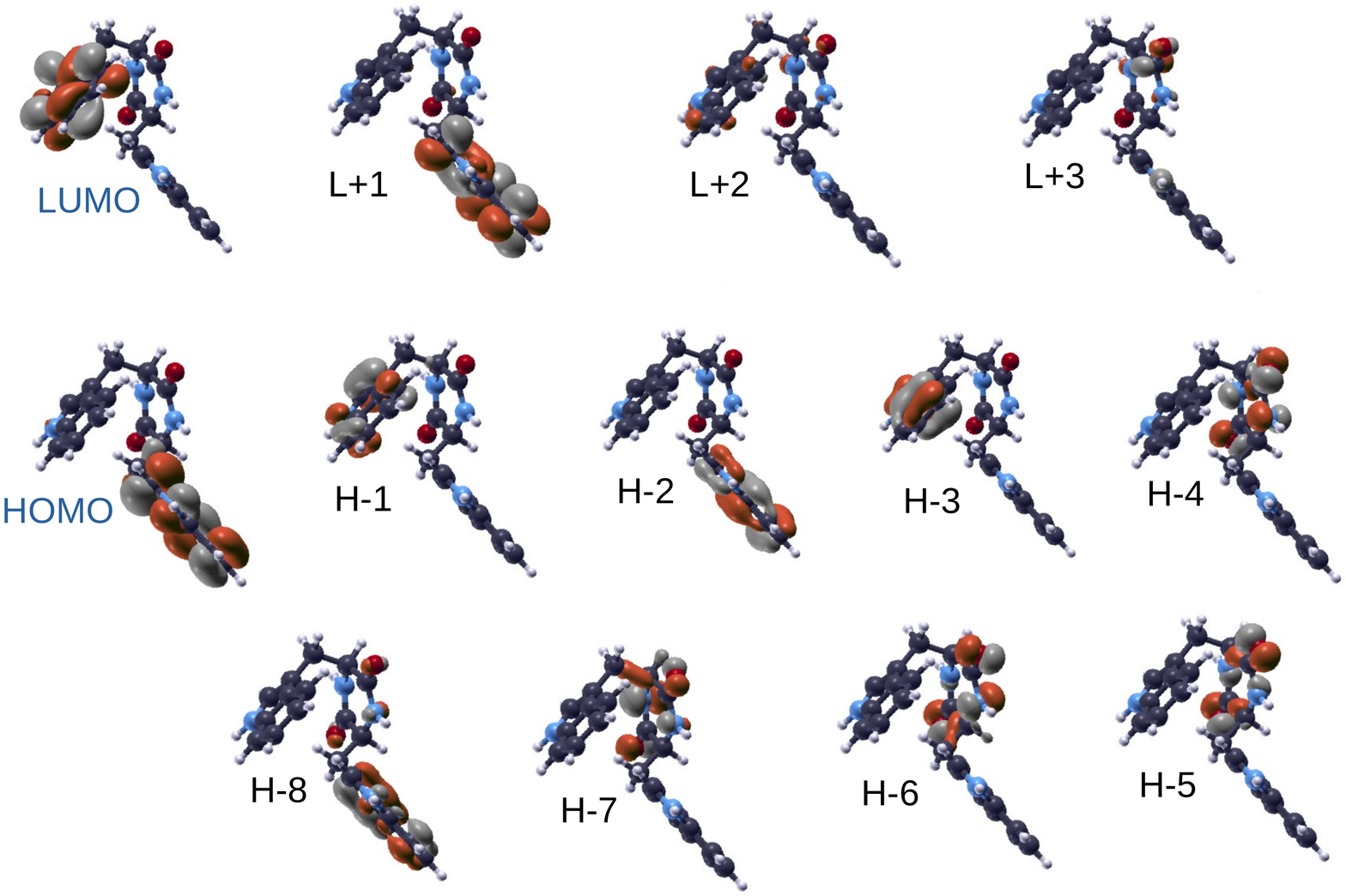}
  \caption{DFT B3LYP orbitals of TrpTrp conformer 3.}
  \label{fig:WF_B3LYP_TrpTrp}
\end{figure}

In Figure~\ref{fig:WF_B3LYP_TrpTrp} we report the spatial localization of QE DFT B3LYP highest occupied and lowest unoccupied electronic states of the TrpTrp lowest energy conformer (conformer 3). Interestingly, molecular orbitals ranging from HOMO-3 to LUMO+1 are alternately localized on the indole ring of either of the two tryptophan amino acids. This is likely due to structural differences responsible for a different chemical environment of the two otherwise identical indole groups. Molecular orbitals localized on the diketopiperazine ring resulting from peptide cyclization appear at lower energies (HOMO-4 to HOMO-7 levels).

\section{Discussion}
The measured photoemission spectra of the three cyclo-dipeptides under study (see Fig.~\ref{fig:3expPES} and black curves in Fig. \ref{fig:levGP}, \ref{fig:levTrpTyr_conf},\ref{fig:levTrpTrp_conf}) share some general characteristics, namely a first feature in the 8 - 10 eV energy region, followed after a gap of about 2 eV by a series of broader bands.  In the first feature two structures are clearly visible in the spectra of TrpTrp and TrpTyr, while they merge in the GlyPhe spectrum. 
The energy of the first feature in the experimental photoelectron spectra increases from 7.97±0.04 eV of TrpTrp to 8.18±0.04 eV in TrpTyr and then 9.54±0.02 eV in GlyPhe.
The trend is the same of the ionization potential of the three aromatic aminoacids Trp, Tyr and Phe, confirming the role of the aromatic chromophore, i.e. the side chain in the aromatic amino acids building the cyclo-dipeptide, in determining the chemical physics properties of the peptide. This is consistent with previous observations in the photoelectron spectra of other cyclo-dipeptides containing amino acids with an aromatic side chain.\cite{Arachchilage_JCP_2012} 
On the other hand, the ionization potentials in the cyclo-dipeptides are higher than those of the constituent aromatic amino acids. This can be considered as a sign of the stability effect induced by the orbitals of the DKP ring on the ones of the aromatic ring of the side chain. 

Indeed, the IE is given by the difference between $E^N_0$ (the ground state total energy with N electrons) and $E^{N-1}_0$ (the ionized state with N-1 electrons), so that, when comparing the IE values of different molecules, the observed trends will depend non-trivially on the combined effect of variations in  $E^N_0$ and in $E^{N-1}_0$.
In particular, the trend observed for the IE values of the three investigated cyclo-dipeptides c-GlyPhe, c-TrpTyr, c-TrpTrp, {\it i.e.} lower IEs for larger aromatic system, can be rationalized by considering that a larger aromatic group will make the ionized molecule more stable (lowering of $E^{N-1}_0$) due to charge delocalization.
When comparing the IEs of these cyclo-dipeptides with those of the corresponding single amino acids, instead, the above-mentioned stabilizing effect of the DKP ring on the neutral molecule (lowering of $E^N_0$) can explain the higher values of IE found for cyclo-dipeptides with respect to single amino acids.

\begin{figure*}[h!]
\centering
 \includegraphics[width=0.9\textwidth]{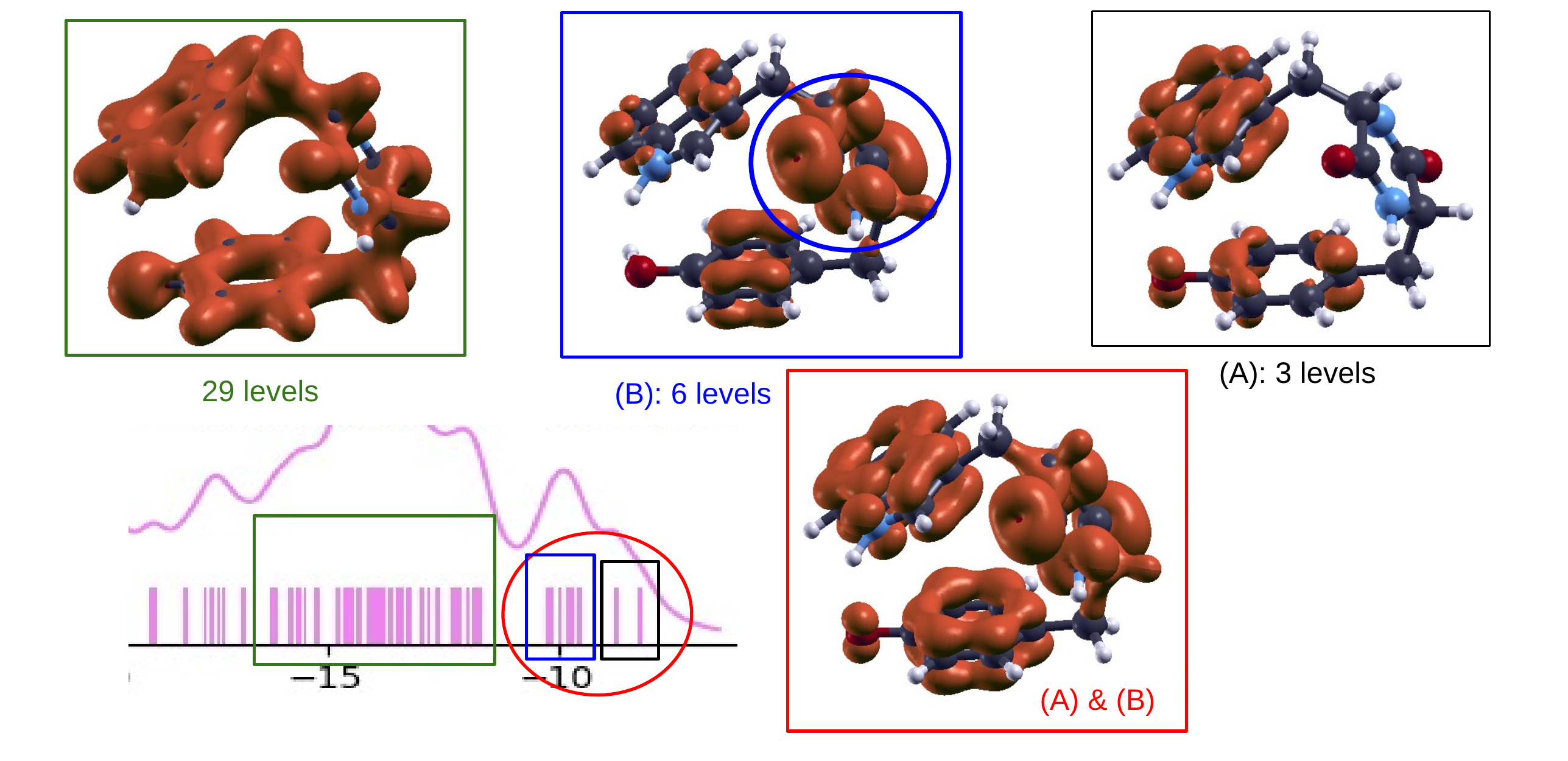}
  \caption{Bottom left: [-20,-5] eV part of the plot (see Fig.~\ref{fig:levTrpTyr_conf} panel (a)) showing the B3LYP calculated DOS of TrpTyr conformer 1. Other panels: plots showing the density of states, for TrpTyr conformer 1, integrated on selected energy ranges, corresponding to specific regions of the DOS plot.}
  \label{fig:ildos}
\end{figure*}

Similar trends are predicted by B3LYP and CCSD calculated densities of electronic states (colored curves in Fig. \ref{fig:levGP}, \ref{fig:levTrpTyr_conf},\ref{fig:levTrpTrp_conf}). From the numerical simulations we can analyze the orbitals which determine the ionization potential and contribute to the first peak in the DOS, in the energy range between 8-10 eV. To this end we plot in Fig.\ref{fig:ildos}, for the case of the TrpTyr, the electronic density obtained from the orbitals in the selected energy range. Similar informations can be obtained for GlyPhe and TrpTrp, by looking at the orbitals in Figures  \ref{fig:WF_B3LYP_GP1} and ~\ref{fig:WF_B3LYP_TrpTrp}. From the plots we clearly see that the ionization potential is determined by “pz-like" orbitals in the aromatic rings and that these orbitals provide a large contribution to the whole first feature (consisting of two substructures, which merge for GlyPhe only) in the DOS. The second substructure in this first DOS feature shows contributions also from the orbitals on the DKP ring (highlighted in blue in Fig.\ref{fig:ildos}). In the case of phenylalanine, the energy of the ionization of the phenyl ring approaches the one of the DKP ring~\cite{Arachchilage_JCP_2010}, and the DOS shows a set of six states very close to each other. As a consequence the highest occupied orbitals of the phenyl side chain are mixed with the DKP ones, and produce the single broader feature observed in the experiment as well as in the calculated DOS for GlyPhe. After these frontier orbitals, theory predicts a region void of states consistently with the gap observed in the experimental spectra.  Then a region with a high density of states (29 in the case of TrpTyr as shown in Fig.\ref{fig:ildos}) follows. These electronic states lie in the planes of the aromatic rings, primarily due to the network of sp$^2$ orbitals of the carbon atoms in these rings.

All the above considerations are based on the analysis of the B3LYP orbitals, which is simple to perform. The good quality of the B3LYP description is confirmed by the agreement with the experimental data, but also by the more refined calculations using the CCSD approach. The latter confirm that the signatures of PES are mostly related to single particle features, with only a very weak satellite appearing in GlyPhe, {\it i.e.} the smallest of the molecules. This is in contrast to what often happens in small molecules where mixing with double and higher order excitations can have an important role. The theoretical description is reminiscent of what happens in extended systems which involve mostly $s$ and $p$ electrons. In such cases DFT provides a reasonable description of the electronic density, and only the eigenvalues need to be corrected. Indeed, for GlyPhe only, we also performed calculations within the state of the art DFT+GW scheme, usually employed to describe extended systems, and found good agreement with the CCSD scheme. The key ingredient of the GW scheme is the screening, whose role grows in importance with the system size, while making other correlation effects included in schemes like CCSD less important. 
GW is not on top of the CCSD results, but it is reasonably close, and it also confirms that the PES can be well described in terms of single-particle features.

This can be further understood in terms of the available phase space. When the system size is increased, so is the number of single particle orbitals, which then provide enough phase space to reach a good description of the electronic properties of the molecules. Observations related to the system size can be done also for the HOMO-LUMO gap and its trend with the molecule size. The HOMO-LUMO gap provides an indication of the stability of the activated molecule towards further chemical reactions. In our case we found the HOMO-LUMO gap tends to decrease as the size of the side chain group becomes larger. This is similar to what happens for example in solid state physics when moving from smaller to bigger nano-structures, to the bulk limit. The bigger the system, the more available phase space to relax, the smaller the electronic gap.

\section{Conclusions}
\label{sect:concl}
We have performed a combined theoretical and experimental study of three cyclo-dipeptides built on amino acids with an aromatic side chain. Theory has allowed to assign the main features of the photoelectron spectra, which share the characteristic of a feature due to the frontier orbitals located on the aromatic ring.  Both theory and experimental data confirm that the ionization energy decreases with increasing size of the aromatic side chains.
These two observations indicate the role of the side chain in determining the (photo)chemical properties of the molecules. The detailed analysis of the spatial localization of electronic states performed in this work will pave the way to the interpretation of the fragmentation of these cyclo-dipeptides by VUV radiation and on the related topic of the charge mobility following ionization between the functional groups making up the cyclo-dipeptide.

Moreover, we also used the molecules to analyze different computational schemes, in particular comparing an accurate quantum chemistry scheme, based on CCSD, with the state of the art approach for extended systems, GW on top of DFT. The goal was to validate the computationally cheaper DFT results; but also to draw considerations related to the importance of dynamical corrections ({\it i.e.} multiple excitations or satellites), electronic screening, and the system size. We consider CCSD as the reference scheme for these molecules, but we also observe that their size is approaching the point where the DFT+GW scheme can also be considered a good alternative.

\section*{Author Contributions}

DS, and LA designed the project. EM, GM, and PA performed the numerical simulations.
LA, PB, LC, FV, YW, RBV, MN, MS, MV, CA, RR performed the experimental measurements.  EM, DS, GM, and LA wrote the manuscript. All the authors have carefully read the manuscript.

\section*{Conflicts of interest}
There are no conflicts to declare.

\section*{Acknowledgements}
The present work was performed in the framework the PRIN 20173B72NB research project ``Predicting and controlling the fate of biomolecules driven by extreme-ultraviolet radiation'', which involves a combined experimental and theoretical study of electron dynamics in biomolecules with attosecond resolution.



\balance


 \bibliographystyle{rsc} 

\providecommand*{\mcitethebibliography}{\thebibliography}
\csname @ifundefined\endcsname{endmcitethebibliography}
{\let\endmcitethebibliography\endthebibliography}{}

\end{document}


\pagestyle{fancy}
\thispagestyle{plain}
\fancypagestyle{plain}{
\renewcommand{\headrulewidth}{0pt}
}

\makeFNbottom
\makeatletter
\renewcommand\LARGE{\@setfontsize\LARGE{15pt}{17}}
\renewcommand\Large{\@setfontsize\Large{12pt}{14}}
\renewcommand\large{\@setfontsize\large{10pt}{12}}
\renewcommand\footnotesize{\@setfontsize\footnotesize{7pt}{10}}
\makeatother

\renewcommand{\thefootnote}{\fnsymbol{footnote}}
\renewcommand\footnoterule{\vspace*{1pt}%
\color{cream}\hrule width 3.5in height 0.4pt \color{black}\vspace*{5pt}} 
\setcounter{secnumdepth}{5}

\makeatletter 
\renewcommand\@biblabel[1]{#1}            
\renewcommand\@makefntext[1]%
{\noindent\makebox[0pt][r]{\@thefnmark\,}#1}
\makeatother 

\renewcommand{\thefigure}{S\arabic{figure}} 
\renewcommand{\thetable}{S\arabic{table}} 

\sectionfont{\sffamily\Large}
\subsectionfont{\normalsize}
\subsubsectionfont{\bf}
\setstretch{1.125} 
\setlength{\skip\footins}{0.8cm}
\setlength{\footnotesep}{0.25cm}
\setlength{\jot}{10pt}
\titlespacing*{\section}{0pt}{4pt}{4pt}
\titlespacing*{\subsection}{0pt}{15pt}{1pt}

\fancyfoot{}
\fancyfoot[LO,RE]{\vspace{-7.1pt}\includegraphics[height=9pt]{head_foot/LF}}
\fancyfoot[CO]{\vspace{-7.1pt}\hspace{11.9cm}\includegraphics{head_foot/RF}}
\fancyfoot[CE]{\vspace{-7.2pt}\hspace{-13.2cm}\includegraphics{head_foot/RF}}
\fancyfoot[RO]{\footnotesize{\sffamily{1--\pageref{LastPage} ~\textbar  \hspace{2pt}\thepage}}}
\fancyfoot[LE]{\footnotesize{\sffamily{\thepage~\textbar\hspace{4.65cm} 1--\pageref{LastPage}}}}
\fancyhead{}
\renewcommand{\headrulewidth}{0pt} 
\renewcommand{\footrulewidth}{0pt}
\setlength{\arrayrulewidth}{1pt}
\setlength{\columnsep}{6.5mm}
\setlength\bibsep{1pt}

\makeatletter 
\newlength{\figrulesep} 
\setlength{\figrulesep}{0.5\textfloatsep} 

\newcommand{\topfigrule}{\vspace*{-1pt}%
\noindent{\color{cream}\rule[-\figrulesep]{\columnwidth}{1.5pt}} }

\newcommand{\botfigrule}{\vspace*{-2pt}%
\noindent{\color{cream}\rule[\figrulesep]{\columnwidth}{1.5pt}} }

\newcommand{\dblfigrule}{\vspace*{-1pt}%
\noindent{\color{cream}\rule[-\figrulesep]{\textwidth}{1.5pt}} }

\makeatother

 \begin{@twocolumnfalse}
{\includegraphics[height=30pt]{head_foot/PCCP}\hfill\raisebox{0pt}[0pt][0pt]{\includegraphics[height=55pt]{head_foot/RSC_LOGO_CMYK}}\\[1ex]
\includegraphics[width=18.5cm]{head_foot/header_bar}}\par
\vspace{1em}
\sffamily
\begin{tabular}{m{4.5cm} p{13.5cm} }

\includegraphics{head_foot/DOI} & \noindent\LARGE{\textbf{Supplementary Information: A systematic study of the valence electronic structure of cyclo(Gly-Phe), cyclo(Trp-Tyr) and cyclo(Trp-Trp) dipeptides in gas phase$^\dag$} } \\
\vspace{0.3cm} & \vspace{0.3cm} \\

 & \noindent\large{Elena Molteni,$^{\ast}$\textit{$^{a,b}$} Giuseppe Mattioli,\textit{$^{a}$} Paola Alippi,\textit{$^{a}$} Lorenzo Avaldi,\textit{$^{a}$} Paola Bolognesi,\textit{$^{a}$} Laura Carlini,\textit{$^{a}$} Federico Vismarra,\textit{$^{c1,c2}$} Yingxuan Wu,\textit{$^{c1,c2}$} Rocio Borrego Varillas,\textit{$^{c2}$} Mauro Nisoli,\textit{$^{c1,c2}$} Manjot Singh,\textit{$^{d}$} Mohammadhassan Valadan,\textit{$^{d,e}$} Carlo Altucci,\textit{$^{d,e}$}  Robert Richter\textit{$^{f}$} and Davide Sangalli\textit{$^{a,b}$}} \\

\end{tabular}
\vspace{0.6cm}


\renewcommand*\rmdefault{bch}\normalfont\upshape
\rmfamily
\section*{}
\vspace{-1cm}


\footnotetext{\textit{$^{a}${Istituto di Struttura della Materia-CNR (ISM-CNR), Area della Ricerca di Roma 1, Via Salaria km 29,300, CP 10, Monterotondo Scalo, Roma, Italy; E-mail (EM): elena.molteni@mlib.ism.cnr.it}}}
\footnotetext{\textit{$^{b}${Dipartimento di Fisica, Università degli Studi di Milano, via Celoria 16, I-20133 Milano, Italy}}}
\footnotetext{\textit{$^{c1}${Dipartimento di Fisica, Politecnico di Milano, Piazza Leonardo da Vinci, 32, Milano, Italy}}}
\footnotetext{\textit{$^{c2}${CNR-Istituto di Fotonica e Nanotecnologie, Piazza Leonardo da Vinci, 32, Milano, Italy}}}
\footnotetext{\textit{$^{d}${Dipartimento di Scienze Biomediche Avanzate, Università degli Studi di Napoli Federico II, via Pansini 5, I-80131 Napoli, Italy}}}
\footnotetext{\textit{$^{e}${Istituto Nazionale Fisica Nucleare (INFN), Sezione di Napoli, Napoli, Italy} }} 
\footnotetext{\textit{$^{f}${Sincrotrone Trieste, Area Science Park, Basovizza, Trieste, Italy}}}

\section{Unbound states in the continuum: reciprocal space vs localized basis-set}

As shown in Figure~\ref{fig:GP1_DOS_occ_empty} for c-GlyPhe conformer 1, delocalized unbound continuum states can be captured when using plane waves basis sets, i.e in our case in the QuantumESPRESSO (QE) Hartree-Fock (yellow) and B3LYP (magenta) datasets, while approaches using localized basis sets, such as our CCSD dataset starting from a ORCA HF calculation can only capture ``resonances'', i.e. unbound states which are (partially) localized on the molecule. Indeed our two QuantumESPRESSO datasets show a continuum-like distribution of states with positive energy, yielding a huge increase in the density of states with respect to the negative energy (bound states) region, while in the CCSD dataset the unbound states, although more dense than the bound ones, are not able to mimick a continuum distribution, and therefore the DOS of unbound states does not show a relevant increase with respect to that of bound states. The GW dataset cannot be taken into account from this point of view, since we have performed this calculation for few QuantumESPRESSO B3LYP empty levels only.

\begin{figure}[th]
\centering
 \includegraphics[width=0.8\textwidth]{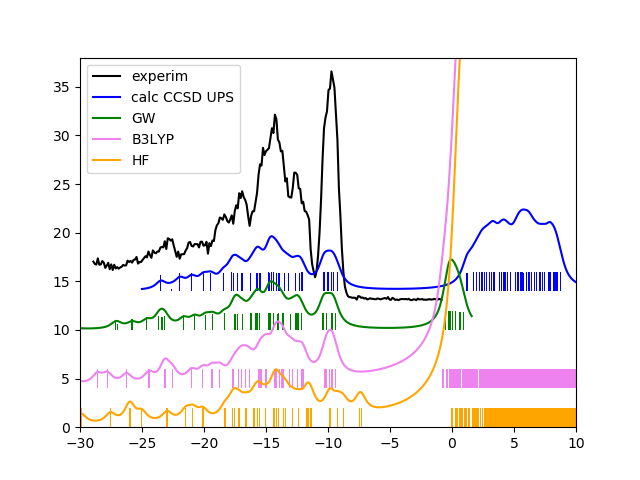}
 \caption{c-GlyPhe conformer 1. QuantumESPRESSO Hartree-Fock (yellow) and B3LYP (magenta) DOS for the occupied and empty electronic levels, compared with the GW (green), and CCSD (blue) PE spectral function and with experimental PE spectrum (black). GW corrections are reported for the first eleven QuantumESPRESSO B3LYP empty levels only. 
The vacuum level is set at 0 energy. The EOM-CCSD calculation has been performed starting from the HF calculation with the localized basis ORCA code. In the HF, B3LYP and GW cases occupied levels are shifted by +2 eV, -2.5 eV, -1.47 eV respectively, with empty levels unshifted.}
 \label{fig:GP1_DOS_occ_empty}
\end{figure}

In Figure~\ref{fig:GP1_WFloc_empty} we show an example of ``resonance``: the state (red tick) at E $\approx$ 2.3 eV, obtained within Hartree-Fock using the localized basis ORCA code, in the highlighted region in the plot of unbound states for c-GlyPhe conformer 1, lies at approximately the same energy as two unbound states (yellow ticks) obtained with HF using the plane wave QE code. In the top and bottom left panels we show the spatial localization of these latter two states, which indeed are rather delocalized, but they also show some density on the molecule, at a difference with e.g. the QE HF LUMO (bottom right panel) - lying at en energy value (not shown here) with no ORCA HF states - which is definitely delocalized, without any density on the molecule.

\begin{figure}[h]
\centering
\includegraphics[width=0.8\textwidth]{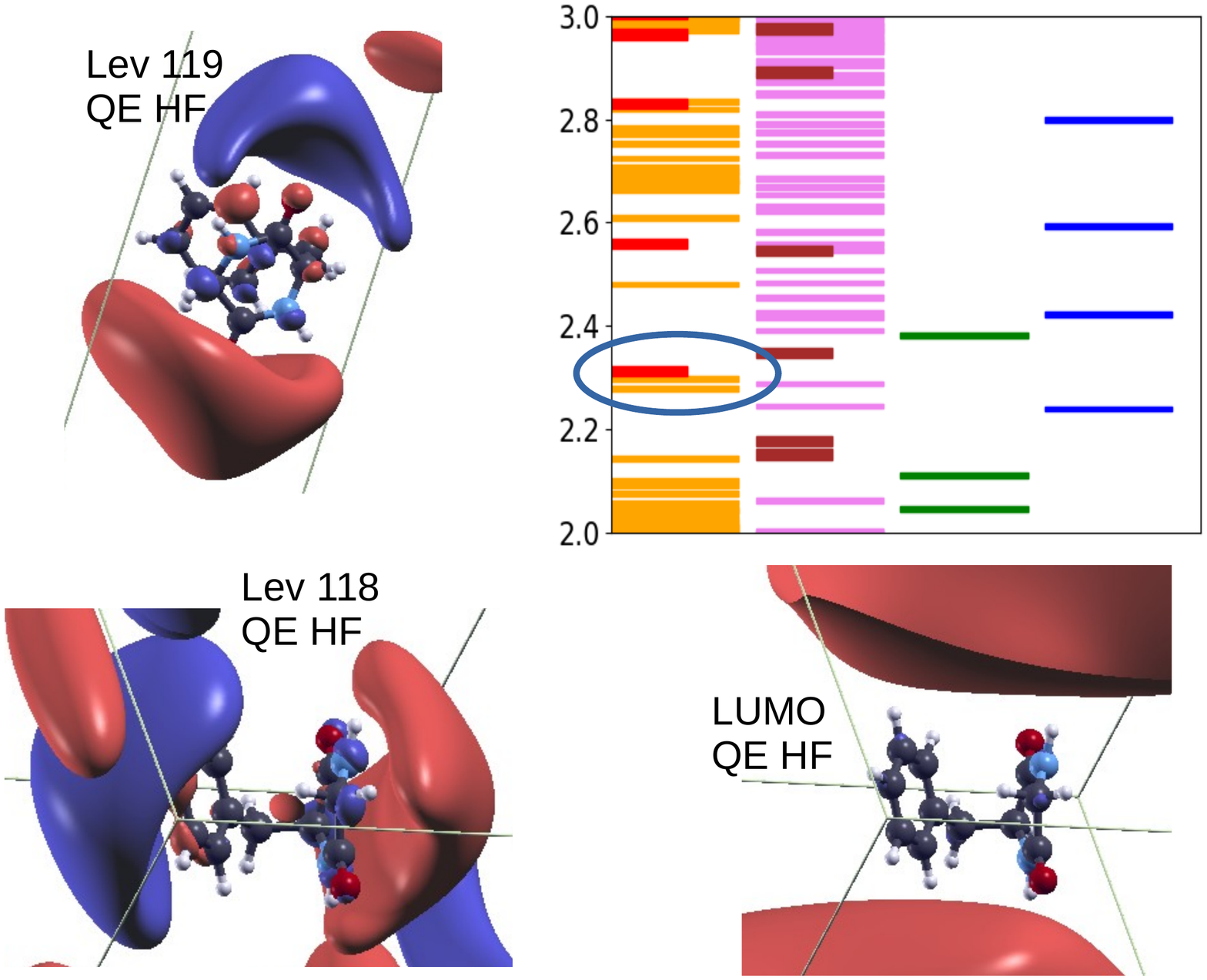}
\caption{c-GlyPhe conformer 1. Top right panel: QuantumESPRESSO Hartree-Fock (yellow), Orca Hartree-Fock (red), QuantumESPRESSO B3LYP (magenta), Orca B3LYP (brown) unoccupied electronic energy levels compared with GW (green) and CCSD (blue) inverse PE energies. GW corrections are reported for the five bound QuantumESPRESSO B3LYP empty levels only. The vacuum level is set at 0 energy. Top left and bottom left panels: spatial localization of the two QuantumESPRESSO Hartree-Fock orbitals highlighted in the top right panel, with energies corresponding to a ``resonance'' i.e. an empty state partially localized on the molecule, obtained by a HF calculation with the Orca localized basis code.  Bottom right panel: spatial localization of the QuantumESPRESSO Hartree-Fock LUMO level.}
\label{fig:GP1_WFloc_empty}
\end{figure}

\clearpage

\section{Energy ordering of different conformers}

In Table~\ref{tab:Eord_cellsize} we show results on the dependence of the plane-wave QuantumESPRESSO DFT B3LYP total energy of the five lowest energy conformers of the c-TrpTrp dipeptide on cell size, using a simple cubic (sc) cell. Conformers are labeled (conf1, conf2 etc) according to their energy ordering found in the initial tight-binding conformational search. Upon increasing the cell side length from 40 a.u. (the one used for all three dipeptides throughout the present work) up to 70 a.u., the energy ordering switches from 3, 2, 1, 4, 5 to 3, 1, 2, 4, 5, and the energy difference among conformations 1, 2 and 3 decreases, so that the situation tends to approach that found in the initial tight-binding conformational search (where the ordering was 1, 2, 3, 4, 5). 
In all these cases, the five analyzed conformers can be separated, according to their total energies, into a first group (conformers 1, 2 and 3) always within a maximum energy range of 0.013 eV from each other, and a second group (conformers 4 and 5) differing in energy by more than 0.041 eV from the highest energy conformer of the first group.

\begin{table}[h]
 \begin{tabular}{||c|c|c|c|c||} 
 \hline
        &   40 a.u.  &  50 a.u.   &  60 a.u.   & 70 a.u.       \\
        [0.1ex] 
 \hline\hline
  conf1 &   -417.0847318 Ry  &   -417.0848135 Ry   &  -417.0850968 Ry &  -417.0851789 Ry          \\ 
        &     +0.011 eV      &     +0.013 eV       &    +0.004 eV     &    +0.003 eV          \\
 \hline
  conf2 &   -417.0850402 Ry  &   -417.0849780 Ry   &  -417.0849742 Ry &  -417.0850295 Ry       \\ 
        &     +0.007 eV      &     +0.011 eV       &    +0.006 eV     &    +0.005 eV            \\
 \hline
  conf3 &   -417.0855624 Ry  &  -417.0857903 Ry    &  -417.0853841 Ry &  -417.0853827 Ry        \\ 
        &      0 eV          &     0 eV            &     0 eV         &     0 eV                \\
 \hline
  conf4 &   -417.0812527 Ry  &  -417.0812544 Ry    &  -417.0813554 Ry &  -417.0814165 Ry       \\ 
        &     +0.059 eV      &    +0.062 eV        &    +0.055 eV     &    +0.054 eV           \\
 \hline
  conf5 &   -417.0747257 Ry  &  -417.0753195 Ry    &  -417.0752128 Ry &  -417.0752561 Ry       \\ 
        &     +0.147 eV      &    +0.142 eV        &    +0.138 eV     &    +0.138 eV            \\
 \hline
\end{tabular}
\caption{c-TrpTrp dipeptide. QuantumESPRESSO DFT B3LYP total energies and energy differences with respect to the lowest energy conformer, of the five lowest energy conformers from the tight binding conformational search: dependence on the cell size, using a simple cubic (sc) cell.}
\label{tab:Eord_cellsize}
\end{table}

Still within QuantumESPRESSO DFT geometry optimizations, we also find the 3, 1, 2, 4, 5 energy ordering if we switch from the B3LYP exchange-correlation functional with NC Martins-Troullier pseudopotentials to the PBE0 one with ONCV pseudopotentials, both with the D3 dispersion correction (first and second column in Table~\ref{tab:Eord_QEfunct}). Increasing the kinetic energy cutoff for the exact exchange operator from 90 Ry to 135 Ry (first and third column) within B3LYP+D3 yields instead no remarkable difference, and the energy ordering remains 3, 2, 1, 4, 5. Upon using a stricter convergence criterion on forces in the geometry relaxation (fourth vs. first column) the energy ordering of conformers becomes 3, 1, 2, 4, 5, with a very small energy difference (0.002 eV) between conformers 1 and 3. By combining the use of a larger cell (70 a.u.) with the above-mentioned strict convergence criterion, we obtain the following energy ordering: 1, 3, 2, 4, 5, with 0.001 eV only between conformers 1 and 3 (fifth column). 
Also in all the calculations reported in this Table, the energy difference between the group of the three lowest energy conformers (1,2,3) and the following conformers (4, 5), is larger than those among conformers 1, 2 and 3.

\begin{table}[h]
 \begin{tabular}{||c|c|c|c|c|c||} 
 \hline
        &   B3LYP+D3            &  PBE0+D3           &  B3LYP+D3                  &  B3LYP+D3                   &  B3LYP+D3                \\
        &   sc, 40 a.u.   &  sc, 40 a.u. &  fock, sc, 40 a.u.   &  tight, sc, 40 a.u.   & tight, sc, 70 a.u. \\
        [0.1ex] 
 \hline\hline
  conf1 & -417.0847318 Ry  &  -417.7950519 Ry   &  -417.0848732 Ry   &  -417.0865420 Ry   &  -417.08657453 Ry   \\ 
        &   +0.011 eV      &    +0.017 eV       &    +0.012 eV       &    +0.002 eV       &     0 eV            \\
 \hline 
  conf2 & -417.0850402 Ry  &  -417.7946484 Ry   &  -417.0851763 Ry   &  -417.0852177 Ry   &  -417.08530383 Ry   \\ 
        &   +0.007 eV      &    +0.022 eV       &    +0.008 eV       &    +0.020 eV       &    +0.017 eV        \\
 \hline 
  conf3 & -417.0855624 Ry  &  -417.7962707 Ry   &  -417.0857429 Ry   &  -417.0867170 Ry   &  -417.08646838 Ry   \\ 
        &    0  eV         &     0 eV           &     0 eV           &     0 eV           &    +0.001  eV       \\
 \hline 
  conf4 & -417.0812527 Ry  &  -417.7924806 Ry   &  -417.0813965 Ry   &  -417.0829814 Ry   &  -417.08292751 Ry   \\ 
        &   +0.059 eV      &    +0.052 eV       &    +0.059 eV       &    +0.051 eV       &    +0.050 eV        \\
 \hline 
  conf5 & -417.0747257 Ry  &  -417.7862192 Ry   &  -417.0748752 Ry   &  -417.0759140 Ry   &  -417.07570804 Ry   \\ 
        &   +0.147  eV     &    +0.137 eV       &    +0.148 eV       &    +0.147 eV       &    +0.148  eV       \\
 \hline 
\end{tabular}
\caption{c-TrpTrp dipeptide. QuantumESPRESSO DFT total energies and energy differences with respect to the lowest energy conformer, of the five lowest energy conformers from the tight binding conformational search: dependence on: type of exchange-correlation functional (2nd column: PBE0 with ONCV pseudopotentials), kinetic energy cutoff for the exact exchange operator (3rd column: increased from 90 Ry to 135 Ry), convergence criterion on forces in the geometry relaxation (4th column: decreased from 5.0$^{-4}$ Ha/bohr to 5.0$^{-5}$ Ha/bohr), cell size (5th column). In all calculations we used a simple cubic (sc) cell.}
\label{tab:Eord_QEfunct}
\end{table}

In Table~\ref{tab:Eord_ORCAfunct} we report results obtained with the ORCA code, using localized basis sets: calculations were performed  either within DFT, varying the treatment of the vdW interactions and the exchange-correlation functional (columns 1 to 3), or with the configuration interaction CCSD(T) method, known to be highly accurate (column 4). 
In all these calculations the lowest energy geometry is that of conformer 1, and the energy ordering of conformers is either 1, 2, 3, 4, 5 or 1, 3, 2, 4, 5. 
Interestingly, the ORCA DFT calculation with the M06-2X meta-GGA hybrid functional yields the same energy ordering - with similar total energy values - as the much more computationally expensive DPLNO-CCSD(T) method.
For really small total energy differences such as those found in some of the calculations analyzed so far, differences in zero point energies (ZPE, neglected so far) may alter the energy ordering of the conformers.
In column 5 we report the zero point energy and entropy corrections to the ORCA DFT B3LYP-D4 total energies, using the def2-TZVPP basis set: these corrections have been applied to the DPLNO-CCSD(T) energies, and the results with the energy difference with respect to the most stable conformer are listed as third and fourth item in each of the cells in the fourth column. The ZPE-corrected CCSD(T) energies yield a 3, 1, 2, 4, 5 conformer ordering, with however 0.0004 eV only between conformers 1 and 3. 

\begin{table}[h]
 \begin{tabular}{||c|c|c|c|c|c||} 
 \hline
        &   B3LYP+D3       &   B3LYP+D4        &    M062X+D3        &    DLPNO-CCSD(T)     &   ZPE-TS (B3LYP)    \\
             [0.1ex] 
 \hline\hline  
  conf1 & -1219.828587 Ha  &  -1219.826531 Ha  &  -1219.9846926 Ha  &  -1218.138127763 Ha  &  0.34075797 Ha  \\
        &     0 eV         &      0 eV         &      0 eV          &      0 eV            &    \\
        &                  &                   &                    &  -1217.797369793 Ha  &    \\
        &                  &                   &                    &     +0.0004 eV       &    \\
 \hline       
 conf2  & -1219.827762 Ha  &  -1219.825807 Ha  &  -1219.9833174 Ha  &  -1218.136843354 Ha  &  0.34093320 Ha  \\
        &    +0.022 eV     &     +0.020 eV     &     +0.037 eV      &     +0.035 eV        &    \\
        &                  &                   &                    &  -1217.795910154 Ha  &    \\
        &                  &                   &                    &     +0.040  eV       &    \\       
 \hline
 conf3  & -1219.827257 Ha  &  -1219.824426 Ha  &  -1219.9837206 Ha  &  -1218.137427990 Ha  &  0.34004460 Ha  \\
        &    +0.036 eV     &     +0.057 eV     &     +0.026 eV      &     +0.019 eV        &    \\
        &                  &                   &                    &  -1217.79738339 Ha   &    \\
        &                  &                   &                    &      0 eV            &    \\
 \hline
 conf4  & -1219.825633 Ha  &  -1219.822603 Ha  &  -1219.9825641 Ha  &  -1218.135709522 Ha  &  0.34026338 Ha  \\
        &    +0.080 eV     &     +0.107 eV     &     +0.058 eV      &     +0.066 eV        &    \\
        &                  &                   &                    &  -1217.79544614 Ha   &    \\
        &                  &                   &                    &     +0.052 eV        &    \\ 
 \hline
 conf5  & -1219.822833 Ha  &  -1219.820356 Ha &   -1219.9781133 Ha &   -1218.132405190 Ha  &  0.34017099 Ha  \\ 
        &    +0.157 eV     &     +0.168 eV    &      +0.179 eV     &      +0.156 eV        &    \\
        &                  &                  &                    &   -1217.79223420 Ha   &    \\
        &                  &                  &                    &      +0.140 eV        &    \\
 \hline
\end{tabular}
\caption{c-TrpTrp dipeptide. Total energies of the five lowest energy conformers from the tight binding conformational search, and energy differences with respect to the lowest energy conformer, obtained with the following methods: Columns 1-3: ORCA DFT with def2-QZVPP basis set: dependence on the type of vdW treatment (2nd column: D4 instead of D3) and on the type of exchange-correlation functional (3rd column: M06-2X meta-GGA hybrid functional); Column 4: ORCA DPLNO-CCSD(T) (without re-optimizing the geometries obtained with ORCA B3LYP-D4); Column 5: zero point energy and entropy corrections to the ORCA B3LYP-D4, def2-TZVPP basis set, energy values.}
\label{tab:Eord_ORCAfunct}
\end{table}

In Table ~\ref{tab:Eord_diff_xTB} we report the conformer energy ordering found in the tight-binding conformational search using different tight-binding Hamiltonians and pairwise dispersion correction treatments, and the effects of including ZPE corrections. Although in the ZPE-uncorrected datasets (columns 1 and 2) conformer 1 is the lowest energy geometry, also here, as in the ORCA DPLNO-CCSD(T) dataset (fourth column of Table~\ref{tab:Eord_ORCAfunct}) the inclusion of the ZPE correction tends to lower the energy of conformer 3. In particular, adding the ZPE correction to the tight-binding GFN2(D4) dataset, conformer 3 becomes the lowest energy geometry of c-TrpTrp. 

\begin{table}[h]
 \begin{tabular}{||c|c|c|c|c||} 
 \hline
        &   GFN2(D4)     &   GFN1(D3)     &   GFN2(D4)+ZPE-TS       &   GFN1(D3)+ZPE-TS        \\
             [0.1ex] 
 \hline\hline  
 conf1  &   -77.21030089 Ha     &   -79.54047362 Ha     &   -76.88384630 Ha       &   -79.21474531 Ha        \\ 
        &     0 eV              &     0 eV              &    +0.005 eV            &     0 eV                 \\
 \hline
 conf2  &   -77.20978280 Ha     &   -79.53861641 Ha     &   -76.88301986 Ha       &   -79.21285896 Ha        \\ 
        &    +0.014 eV          &    +0.051 eV          &    +0.027 eV            &    +0.051 eV             \\
 \hline
 conf3  &   -77.20909176 Ha     &   -79.53741587 Ha     &   -76.88401709 Ha       &   -79.21342276 Ha        \\ 
        &    +0.033 eV          &    +0.083 eV          &     0 eV                &    +0.036 eV             \\
 \hline 
 conf4  &   -77.20879438 Ha     &   -79.53792224 Ha     &   -76.88337462 Ha       &   -79.21344391 Ha        \\ 
        &    +0.041 eV          &    +0.069 eV          &    +0.017 eV            &    +0.035 eV             \\
 \hline 
 conf5  &   -77.20777333 Ha     &   -79.53755072 Ha     &   -76.88199196 Ha       &   -79.21257408 Ha        \\ 
        &    +0.069 eV          &    +0.080 eV          &    +0.055 eV            &    +0.059 eV             \\
 \hline 
 \end{tabular}
\caption{c-TrpTrp dipeptide. Total energies of the five lowest energy conformers from a tight binding conformational search, and energy differences with respect to the lowest energy conformer, obtained using: column 1 (column 2) the GFN2-xTB (GFN1-xTB) tight-binding Hamiltonian with the D4 (D3) pairwise dispersion correction; columns 3 and 4: same two methods, adding the zero point energy and entropy corrections.}
\label{tab:Eord_diff_xTB}
\end{table}

Summarizing, from this detailed analysis of the energy ordering of c-TrpTrp geometries (Tables~\ref{tab:Eord_cellsize} to  ~\ref{tab:Eord_diff_xTB} ), we can conclude that, as for the starting tight-binding conformational search, both the implementations we have tried within the CREST code, i.e. the one with the GFN1 tight-binding Hamiltonian and the D3 pairwise dispersion correction, and the one with GFN2 and D4 (expected to better describe electrostatic and vdW interactions), identify the same five lowest energy conformers.

When starting from the five lowest energy geometries found with the tight-binding conformational search, both DFT and CCSD calculations, with various choices/combinations of xc functionals, basis type, vdW treatment, yield a first ``cluster'' of lowest energy geometries (conformers 1, 2, and 3, with possibly varying mutual ordering) followed, at higher energies, by conformers 4 and 5. Often the total energies of conformers 1, 2 and 3 span a smaller energy range if compared to their energy separation from conformers 4 and 5.  

In particular, within the first cluster of conformers, plane-wave QE DFT calculations (ORCA DFT calculations with localized basis sets) tend to favour conformer 3 (conformer 1).  Within the very accurate - and computationally expensive - CCSD(T) method with the addition of zero point energy corrections, the most stable geometry is conformer 3, although at 0.0004 eV only from conformer 1.

These findings lead us to focus on conformers 1, 2 and 3 of c-TrpTrp as reliable candidates for lowest energy geometries of this dipeptide, on which we then calculated electronic properties.

\clearpage

\section{Spatial localization of EOM-CCSD and of B3LYP orbitals}

\begin{figure}[h]
\centering
 \includegraphics[width=0.6\textwidth]{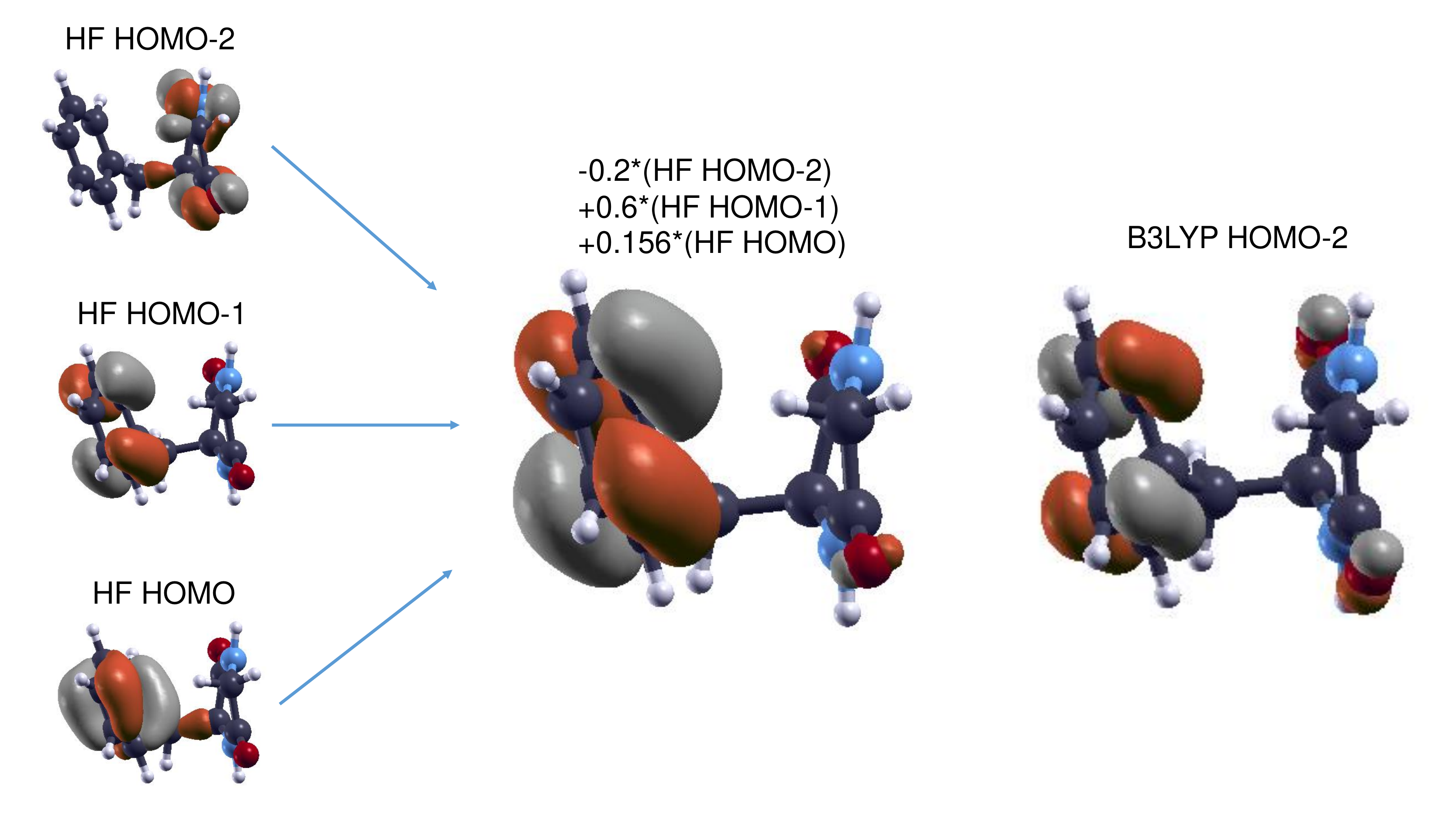}
 \caption{Hartree-Fock orbitals (left)
contributing to the CCSD "IE-1" ionization energy of c-GlyPhe (center), and the B3LYP HOMO-2 (right).  }
 \label{fig:GP1_WF_CCSD_IE-1}
\end{figure}

\begin{figure}[h]
\centering
 \includegraphics[width=0.6\textwidth]{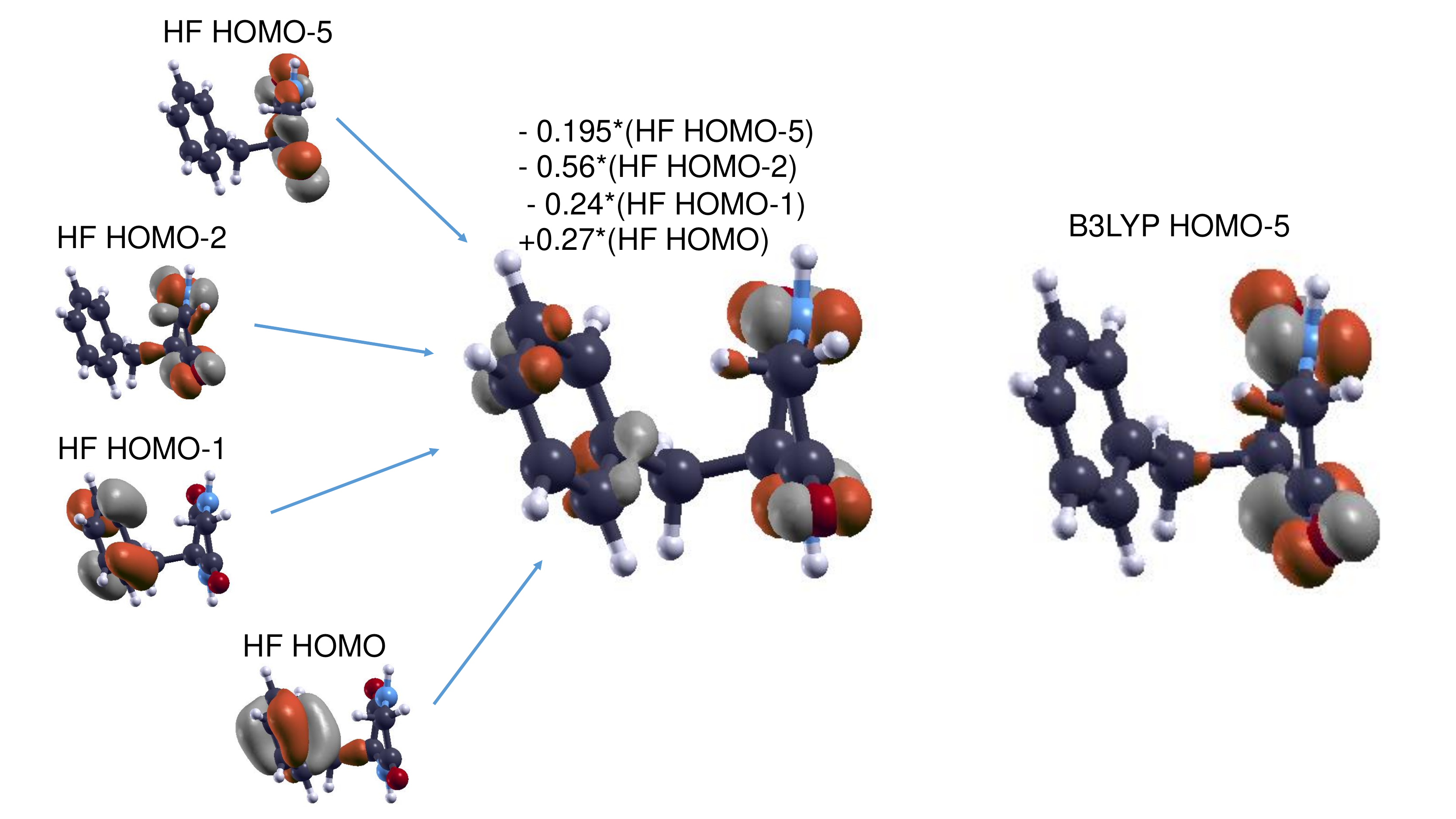}
 \caption{Hartree-Fock orbitals (left)
contributing to the CCSD "IE-2" ionization energy of c-GlyPhe (center), and the B3LYP HOMO-5 (right).  }
 \label{fig:GP1_WF_CCSD_IE-1}
\end{figure}

\clearpage

\section{  Calculated adiabatic ionization energies and electron affinities for the cyclo-dipeptides under study}

\begin{table}[h]
 \begin{tabular}{||c|c|c|c|c||} 
 \hline
  eV & r$^2$-SCAN-3c & B3LYP-D3 & M062X-D3 & DLPNO-CCSD(T)                 \\
 \hline\hline
\multicolumn{5}{|c|}{c-GlyPhe}     \\
 \hline
  IP & 8.15 & 8.35 & 8.93 & 9.11  \\
 \hline
  EA & 0.37 & 0.48 & 0.50 & 0.74  \\
 \hline
\multicolumn{5}{|c|}{c-TrpTyr}     \\  
 \hline
  IP & 6.94 & 7.01 & 7.36 & -     \\
 \hline
  EA & 0.06 & 0.17 & 0.15 & -     \\
 \hline
\multicolumn{5}{|c|}{c-TrpTrp}     \\ 
 \hline 
  IP & 6.82 & 6.94 & 7.40 & -     \\
 \hline
  EA & 0.01 & 0.15 & 0.20 & -     \\
 \hline 
 \end{tabular}
\caption{Adiabatic ionization energies (IE) and electron affinities (EA) for the investigated cyclo-dipeptides, calculated with the meta-GGA exchange-correlation functional r$^2$-SCAN-3c and the mTZVPP basis set (column 1), the global hybrid B3LYP functional (20 $\%$ exact exchange (EXX)) (column 2) and the meta-GGA global hybrid M062X functional (54$\%$ EXX) (column 3), both using the D3 dispersion correction and the def2-QZVPP basis set, and finally with the DLPNO-CCSD(T) method (column 4).
In each of the three cases reported in columns 1, 2 and 3 the geometry has been optimized with the chosen functional; in the DLPNO-CCSD(T) case, due to the large computational weight, IE and EA have been calculated on the B3LYP-D3 optimized geometry.
In all cases the thermochemical contribution (ZPE-TS@298 $^{\circ}$~K) calculated with the r$^2$-SCAN-3c functional has been added.}
\label{tab:adiab_IE}
\end{table}

\begin{table}[h]
 \begin{tabular}{||c|c|c|c|c||} 
 \hline
eV  & B3LYP-D3 adiab  & B3LYP-D3 vert  & M062X-D3 adiab  &  M062X-D3 vert    \\
\hline\hline
\multicolumn{5}{|c|}{c-GlyPhe}     \\
 \hline
 IP & 8.43 & 8.57 & 9.01 & 9.25    \\
 \hline
 EA & 0.68 & 0.81 & 0.70 & 0.91    \\
 \hline
\multicolumn{5}{|c|}{c-TrpTyr}     \\ 
 \hline
 IP & 7.05 & 7.29 & 7.40 & 7.76    \\
 \hline
 EA & 0.38 & 0.63 & 0.36 & 0.71    \\
 \hline
\multicolumn{5}{|c|}{c-TrpTrp}     \\ 
 \hline
 IP & 6.97 & 7.08 & 7.42 & 7.70    \\
 \hline
 EA & 0.37 & 0.47 & 0.42 & 0.64    \\
 \hline 
 \end{tabular}
\caption{Comparison between adiabatic and vertical values of IE and EA for the three investigated cyclo-dipeptides, all obtained using the def2-QZVPP basis set and without the thermochemical contribution ZPE-TS.}
\label{tab:adiab_vs_vert}
\end{table}

 \end{@twocolumnfalse} 